%% file: mvf.tex
\documentclass[twocolumn,aps,prd,floatfix,preprintnumbers,a4paper,nofootinbib,
superscriptaddress,10pt]{revtex4-1}

\usepackage{epsfig}
\usepackage{graphics}
\usepackage{graphicx}
\usepackage{amsmath,amssymb,mathtools}
\usepackage{amsfonts}
\usepackage{bm,soul}
\usepackage[usenames,dvipsnames]{color}
\usepackage{amssymb}
\usepackage{url}
\usepackage{psfrag}
\usepackage{times}
\usepackage[varg]{txfonts}
\usepackage[colorlinks, pdfborder={0 0 0}]{hyperref}
\usepackage{lineno}
\usepackage{verbatim}
\definecolor{LinkColor}{rgb}{0.75, 0, 0}
\definecolor{CiteColor}{rgb}{0.75, 0, 0}
\definecolor{UrlColor}{rgb}{0, 0, 0.75}
\hypersetup{linkcolor=LinkColor}
\hypersetup{citecolor=CiteColor}
\hypersetup{urlcolor=UrlColor}
\usepackage[utf8]{inputenc}
\usepackage{ulem}
\usepackage{algorithm}
\usepackage{algpseudocode} 
\usepackage{setspace} 

\input{definitions}

\newcommand{\cw}{\tilde{\omega}}

\def\jf{j_f}
\def\mf{M_f}

\def\lmn{_{\ell m n}}
\def\LM{_{\bar{\ell} \bar{m}}}
\def\LMlmn{_{\bar{\ell} \bar{m} \ell m n}}
\def\gmvp#1{greedy-multivariate-polynomial#1
  (\texttt{GMVP}#1)\gdef\gmvp{\texttt{GMVP}}}
\def\gmvr#1{greedy-multivariate-rational#1
  (\texttt{GMVR}#1)\gdef\gmvr{\texttt{GMVR}}}

\newcommand{\lmtitle}[2]{\vspace{-0.80cm} \begin{center} \noindent\rule{0.35\paperwidth}{0.3pt} \end{center} \vspace{-0.3cm}}

\def\CwFitCalibrationRegion{\red{0.995} }

\def\pgreedy{\texttt{PGREEDY}}

\newcommand{\Cardiff}{School of Physics and Astronomy, Cardiff University, Queens Buildings, Cardiff, CF24 3AA, United Kingdom}

\begin{document}


\title{On modeling for Kerr black holes: Basis learning, QNM frequencies, and spherical-spheroidal mixing coefficients }

\author{L. London}
\affiliation{\Cardiff}
\author{E. Fauchon-Jones}
\affiliation{\Cardiff}

\begin{abstract}
  Models of black hole properties play an important role in the ongoing direct detection of gravitational waves from black hole binaries. One important aspect of model based gravitational wave detection, and subsequent estimation of source parameters, are the low level modeling of information related to perturbed Kerr black holes. Here, we present new phenomenological methods to model the analytically understood gravitational wave spectra (quasi-normal mode frequencies), and harmonic structure of Kerr black holes (mixing coefficients between spherical and spheroidal harmonics). In particular, we present a greedy-multivariate-polynomial (GMVP) regression method and greedy-multivariate-rational (GMVR) regression method for the automated modeling of polynomial and rational functions respectively. GMVR is a quasi-linear numerical method for interpolating rational functions. It therefore represents a solution to Runge's phenomenon. GMVP is used to develop a model for QNM frequencies that explicitly enforces consistency with the extremal Kerr limit. GMVR is used to develop a model for harmonic mixing coefficients that extends previous results to dominant multipoles with $\ell \le 5$. Both models are the first of their kind to consider black hole spin to vary between -1 and 1, thus naturally connecting the pro and retrograde modes. We discuss the potential use of these models in current and future gravitational wave signal modeling.
\end{abstract}

\date{\today}

\maketitle

\section{Introduction}
In the coming years, expectations for frequent \gw{} detections of increasing \snr{} are high \cite{TheLIGOScientific:2016pea,Abbott:2016nhf}.
Concurrent with \virgo{}, the \aligo{} detectors will enter their third observing run in approximately early 2019.
During this period, a few to dozens of \bbh{} signals are likely to be detected \cite{Abbott:2016nhf,Abbott2018OS}.
In this context, signal detection and subsequent inference of physical parameters hinges upon efficient models for source properties and dynamics \cite{Abbott:2016wiq}.
Most prominently, there is ongoing interest in efficient and accurate signal models for \bbh{} \imr{} \cite{Blackman:2017dfb, London:2017bcn, Hannam:2013oca}.
As the merger of isolated \bh{s} is expected to result in a perturbed Kerr \bh{}, there is related interest in having computationally efficient models for perturbative parameters, namely those that enable evaluation of the related \rd{} radiation \cite{Berti:2005ys}.
\par In particular, a perturbed Kerr \bh{} (e.g. resulting from \bbh{} merger) will have \gw{} radiation that rings down with characteristic \textit{dimensionless} frequencies,
$
	\cw\lmn = \omega\lmn + i/\tau\lmn
$
, where $\omega\lmn$ is the central frequency of the ringing, and $1/\tau\lmn$ is the damping rate.
These discrete frequencies have associated radial and spatial functions which are \textit{spheroidal} harmonic in nature \cite{Leaver85}.
Together, these harmonic functions and frequencies constitute the \qnm{} solutions to \ee{}.
Specifically, they are the eigen-solutions of the source free linearized \ee{} (i.e. \tk{'s} equations \cite{PhysRevLett.29.1114}) for a perturbed \bh{} with final mass, $\mf$, and dimensionless final spin, $\jf$.
These solutions allow \grad{} from generic perturbations to be well approximated by a spectral (multipolar) sum which combines the complex \qnm{} amplitude, $A\lmn$, with the spin $-2$ weighted -2 spheroidal harmonics, $_{-2}S\lmn$.
\begin{align}
	\label{hrd}
	h &= h_{+} - i \, h_{\times}
	  \\ \nonumber
	  &= \frac{1}{r} \sum\lmn A\lmn \; e^{i\,\cw\lmn t} \; _{-2}S\lmn( \jf \cw\lmn,\theta,\phi)
		\\ \nonumber
		&= \frac{1}{r} \sum\LM h\LM(t) \; _{-2}Y\LM(\theta,\phi) \;  .
\end{align}
\par In the first and second lines of \eqn{hrd}, we relate the observable \gw{} polarizations, $h_+$ and $h_\times$, to the analytically understood morphology of the time domain \rd{} waveform.
Here, the labels $\ell$ and $m$ are polar and azimuthal eigenvalues of \tk{'s} angular equations, where total mass and the speed of light are set to unity (i.e. $M=c=1$).
Note that barred indices, namely $\bar{\ell}$ and $\bar{m}$, refer to \textit{spherical} harmonics of spin weight -2, leaving the unbarred indices to refer to the spheroidals.
In the third line of \eqn{hrd}, we represent $h$ in terms of {spherical} harmonic multipoles.
This latter form is ubiquitous for the development and implementation of \imr{} signal models for \bbh{s}.
\par Towards the development of these models, \eqn{hrd} enters in many incarnations.
In the \eob{} formalism, $h\LM$ is modeled such that, after its peak (near merger), the effective functional form reduces (asymptotically) to \eqn{hrd}'s second line \cite{Cotesta:2018fcv, Buonanno:2000ef,
Bohe:2016gbl,Pan:2013rra, Bohe:2016gbl, Nagar:2018zoe}.
This view currently comes with the added assumption that ${_{-2}}S_{\ell m n} = {_{-2}}Y_{\ell m}$, where ${_{-2}}Y_{\ell m}$ are the spins weighted -2 spherical harmonics.
Only where $\jf \cw\lmn=0$ do $_{-2}S\lmn( 0,\theta,\phi) = {_{-2}}Y_{\ell m}(\theta,\phi)$, which makes equating the spherical and spheroidal harmonics approximate at best for general values of $\jf \cw\lmn$.
The consequences of that approximation, in particular the mixing between spherical and spheroidal harmonics, are discussed in reference \cite{London:2014cma,Berti:2014fga,London:2018gaq,Kelly:2012nd}.
This approximation also applies to the \textit{Phenom} models,
where the frequency domain multipoles, $\tilde{h}\LM(f)$, are constructed such that their high frequency behavior is consistent with \eqn{hrd} in the time domain
\cite{Hannam:2013oca, London:2017bcn, Khan:2015jqa, Schmidt:2014iyl, Mehta:2017jpq,Khan:2018fmp}.
\par Both Phenom and \eob{} approaches either use phenomenological models of remnant \bh{} mass and spin to interpolate over tables of \qnm{} frequencies, or phenomenological models for \qnm{} frequencies are used directly.
Either approach typically incurs less computational cost that the direct numerical calculation of \qnm{} frequencies, which may involve, for example, the solving of continued fraction equations \cite{Leaver85}.
In the case of the higher multipole model \texttt{PhenomHM} and its derivative models, fits for the \qnm{} frequencies are used in the process of mapping $\tilde{h}_{22}(f)$ into other $\tilde{h}\LM(f)$ \cite{London:2017bcn}.
In that setting, it is demonstrated that \qnm{} frequencies are linked to the amplitude and phase of each $h\LM$ in not only \rd{}, but also merger and late inspiral, as is implied by the source's causal connectedness pre and post merger.
\par For models that assist tests of the \nht{} (e.g. \cite{Berti:2005ys,London:2018gaq,Carullo:2018sfu}), and thereby only include precise \rd{s}, the perspective of \eqn{hrd}'s second and third lines are used to write each spherical harmonic multipole moment as
\begin{align}
	\label{hlm}
	h\LM = \frac{1}{r} \sum\lmn \, A\lmn e^{i \cw\lmn t} \sigma\LMlmn
\end{align}
where, the spherical-spheroidal mixing coefficient, $\sigma\LMlmn$, is
\begin{align}
		\label{sigma}
		\sigma\LMlmn = \int_{\Omega} {_{-2}S}\lmn \,  {_{-2}}Y^*\LM \, \mathrm{d} \Omega \; .
\end{align}
In \eqn{sigma}, $*$ denotes complex conjugation, and $\Omega$ is the standard solid angle in spherical polar coordinates.
\par In practice, using \eqn{hlm} is computationally efficient:
Whereas the calculation of each ${_{-2}S}\lmn$ involves a series solution which slowly converges for $\jf$ near unity, the calculation of each ${_{-2}Y}\LM$ is achieved using closed form expressions.
It is therefore efficient to use accurate models for $\sigma\LMlmn$ to avoid convergence issues. These can then be used directly to calculate $h\LM$ via \eqn{hlm}, and thereby the \gw{} polarizations via \eqn{hrd}.
\par In this combined context, it is clear that the modeling of \qnm{} frequencies, $\cw\lmn$, and spherical-spheroidal mixing coefficients, $\sigma\LMlmn$, are relevant for a range of \gw{} signal models.
While models for $\cw\lmn$ and $\sigma\LMlmn$ are present in the literature (e.g. \cite{Berti:2005ys, Berti:2014fga, Cook:2014cta}), there exist minor shortcomings which we wish to address here.
\par For both the \qnm{} frequencies and the spherical-sphoidal mixing coefficients, we present the first models which treat \qnm{s} rotating with and against the rotation of the \bh{} as being a part of a single solution parameterized by dimensionless \bh{} spin ranging from -1 to 1.
This perspective reflects the empirical observation that the remnant spin of \bbh{} mergers smoothly connects regions of positive and negative spin relative to the direction of the initial oribital angular momentum \cite{Husa:2015iqa,London:2018gaq}.
\par For the \qnm{} frequencies, it is well known that for nearly extremal \bh{s} (i.e. $\jf \rightarrow 1$) some of the frequencies have zero-damping (i.e. $\tau\lmn \rightarrow \infty$) \cite{Yang:2012pj,Zimmerman:2015trm}.
In the context of \gw{} \da{}, where source parameters are estimated using routines which sample over the space of all possible \bh{} masses and spins, it is useful to have accurate physical behavior in the extremal limit \cite{TheLIGOScientific:2016wfe}.
Like Ref.~\cite{Cook:2014cta}, we present models for $\cw\lmn$ that explicitly account for zero-damping in the extremal Kerr limit.
The models presented here go further by applying outside of the nearly extremal Kerr regime while also accounting for non-zero-damping in modes such as $(\ell,m,n)=(2,1,0)$ and $(3,2,0)$\cite{Yang:2012pj}.
\par For the modeling of $\sigma\LMlmn$, we note that the models of Ref.~\cite{Berti:2014fga} do not appear to include the \qnm{s} which rotate counter to the \bh{} spin direction (i.e. ``mirror-modes'').
Here these \qnm{} are explicitly modeled on a continuation of the positive spin line to negative spin.
\par In parallel, the methods for modeling $\cw\lmn$ and $\sigma\LMlmn$ have been dispersed: different phenomenological techniques have been used under no coherent framework.
Here we will present linear modeling techniques, namely the \gmvp{} and \gmvr{} algorithms, in which model terms are iteratively learned with no initial guess.
The description of \gmvp{} given here is complementary to similar algorithms used to model \qnm{} excitation amplitudes, $A\lmn$, as present in reference \cite{Carullo:2018sfu,London:2018gaq,London:2014cma}.
As we will discuss, the \gmvr{} algorithm is an iterative approach to the (pseudo) linear modeling of multivariate rational functions, wherein iterations of linear inversions are used to refine the ultimately non-linear model.
\par In the rudimentary form presented here, both \gmvp{} and \gmvr{} are intended for use with low noise data (e.g. the results of analytic calculations), and each employs a reverse (or negative) greedy algorithm to counter over modeling \cite{Field:2011mf,Caudill:2011kv}.
As the underlying process for \gmvp{} and \gmvr{} is stepwise regression, highly correlated basis vectors (i.e. polynomial terms) are handled via an approach we will call \textit{degree tempering}.
It will be demonstrated that these approaches are readily capable of modeling the complex valued $\cw\lmn$ and $\sigma\LMlmn$.
Results suggest that the versions of \gmvp{} and \gmvr{} presented here may apply in instances where training data are approximately noiseless, and an initial guess is difficult to obtain.
Both algorithms are publicly available in \texttt{Python} via Ref. \cite{lionel_london_2018_1402516}.
While this paper's fits for the \qnm{} frequencies and mixing coefficients are presented in \eqns{eq:cw_fit_1}{eq:cw_fit_9} and \eqns{eq:ys22220}{eq:ys55550}  respectively, we encourage the reader to use the fits implemented in Ref.~\cite{lionel_london_2018_1402516}: \texttt{positive.physics.cw181003550} (\qnm{} frequencies) and in \texttt{positive.physics.ysprod181003550} (mixing coefficients).
\par The plan of the paper is as follows.
In section \sec{meth}, we outline the \gmvp{} and \gmvr{} algorithms.
In \sec{results}, we demonstrate the application of each algorithm.
We first consider the application of \gmvp{} to the modeling of \qnm{} frequencies.
We then consider the application of \gmvr{} to the modeling of spherical-spheroidal mixing coefficients.
Quantitative comparisons are made between our models and those presented in Refs.~\cite{Berti:2005ys, Berti:2014fga}.
In \sec{discuss}, we review the effectiveness of \gmvp{} and \gmvr{}, and we discuss current and potential applications for these methods.
\section{Methods}
\label{meth}
Within the topic of regression, linear regression has particular advantages.
Its matrix based formulation can be computationally efficient, and it does not require initial guesses for model parameters.
Perhaps most intriguingly, the formal series expansions of smooth functions support linear and rational models (e.g. Pad\'e approximants) that have application to many datasets.
With this in mind, here, we will develop algorithms for the linear (polynomial and rational) modeling of scalar functions (real or complex) of many variables.
\par If we consider a scalar function, $f$, of $N$ variables sampled in $j$, $\vec{x}_j = \{x_{\alpha j}\}_{\alpha=0}^{N}$, then $f(\vec{x}_j)$ can be represented (possibly inaccurately) as a sum over $K$ linearly independent basis functions, $\phi_k(\vec{x}_j)$:
\begin{align}
  \label{lin1}
  f({\vec{x}}_j) \; = \; \sum_{k=0}^{K} \; \mu_{k} \, \phi_k(\vec{x}_j)\; .
\end{align}
The central player in \eqn{lin1} is the set of basis coefficients $\mu_{k}$.
Typically, one chooses or derives $\phi_{k}(\vec{x}_j)$ to capture inherent features of $f(\vec{x}_j)$.
With $\phi_{k}(\vec{x}_j)$ assumed to be known, the linear representation (namely \ceqn{lin1}) is lastly defined the set of $\mu_k$.
\def\vecmu{\vec{\mu}}
\def\vecf{\vec{f}}
\def\hatU{\hat{U}}
\def\hatP{\hat{P}}
\par From here it is useful to note that \eqn{lin1} has a linear homogeneous matrix form.
In particular, defining ${U}_{jk} = \phi_k(\vec{x}_j)$, $f_j = f(\vec{x}_j)$ and $\vec{f} = \hat{U} \; \vec{\mu}$, implies that
\begin{align}
  \label{eq:linsol1}
  \vecmu = \hatP \; \vecf \; ,
\end{align}
where
\begin{align}
  \label{pinv1}
  \hatP = \left( \hatU^{*} \hatU \right)^{-1}\hatU^{*}
\end{align}
is the \textit{pseudo-inverse} \cite{Moore1920,Penrose:1955} of $\hatU$, which exists if $\hatU^{*} \hatU$ is nonsingular.
Here ``$*$'' denotes the conjugate transpose.
\par Equations (\ref{lin1}), and related discussion through \eqn{pinv1} illustrate the most rudimentary solution to the linear modeling problem.
However, there are many ways to expand upon and refine the solution presented thus far.
In the following subsections we will consider two such approaches.
First we will consider the general polynomial modeling of multivariate scalar functions.
This will encompass the \gmvp{} algorithm.
Second, we will build upon the \gmvp{} approach by considering models of rational functions (polynomials divided by polynomials).
To consider these two approaches in a largely automated way (i.e. where the set of possible basis functions is known, but the select basis functions ultimately used are \textit{learned}), we will make use of the \textit{greedy} algorithm approach \cite{Field:2011mf, GVK022791892, 1978AnSta, Pandit84}.
\subsection{A Generic Greedy Algorithm}
While we most often want a single model for a given dataset (e.g. some approximation of $\vecf$ from numerical calculation or experiment), there are often many more modeling choices than desired.
In particular, if we refer to our set of all possible basis functions as our ``\textit{symbol space}'', then the problem of determining how many, and which basis vectors (i.e. symbols) to use is a problem of combinatoric complexity.
\par A well known method for finding an approximate solution to this problem is the so-called ``greedy'' algorithm (e.g. \cite{Field:2011mf, GVK022791892}):
We will iteratively construct models with increasing number of symbols.
The process begins by finding the single symbol (basis vector) that yields the most accurate model in the sense of minimizing the least-squares error.
That encompasses the first iteration of a process in which we will greedily add symbols to our model.
In each subsequent iteration, remaining symbols are added to the model one at a time, resulting in many trial models, each with its own representation error.
The trial model with smallest representation error is kept for the next greedy iteration.
In this way, a list of optimal model symbols is learned.
This forward greedy process ends when the model accuracy, and/or changes thereof, passes a previously specified threshold.
This rough algorithmic picture is encapsulated by Alg. \ref{alg:pgreedy}.
%
\hspace{1cm}
{\scriptsize
\begin{algorithm}[H]
  \caption{A positive (forward) greedy algorithm, \pgreedy{}. Note that a required input, $\mathcal{A}$, is a function that takes in a list of basis symbols, and outputs an estimator of fit error (e.g. $L^2$ norm). In this setting, $\mathcal{A}$ is assumed to have access to peripheral information, such as the training data.}
  \label{alg:pgreedy}
  \begin{algorithmic}[1]
    \State {\bf Input:} $ \{ \mlam_{bulk} = \text{basis symbols}$, $\mathcal{A} = \text{action}$, $tol=\text{greedy tolerance}\}$
    \vskip 10pt
    \State Define empty list of kept symbols: $\mlam_{kept} = \{\}$
    \State Initialize estimator value and loop boolean: $\epsilon_{last} = \mathrm{inf}$, $done = \text{False}$
    \While{not $done$}
      \State $\epsilon_{min} = \epsilon_{last}$
      \For { $\mlam$ in $\mlam^{bulk}$ }
        \State $\mlam_{trial} = \mlam_{kept} \cup \mlam$ {\hskip0.525in} (add $\mlam$ to $\mlam_{kept}$)
        \State $\epsilon = \mathcal{A}(\mlam_{trial})$ {\hskip0.70in} (action returns fit error)
        \If { $\epsilon < \epsilon_{min}$ }
          \State $\epsilon_{min} = \epsilon$  {\hskip0.825in} (store trial min)
          \State $\mlam_{min} = \mlam_{trial}$
        \EndIf
      \EndFor
      \State $done = |\epsilon_{min}-\epsilon_{last}|<tol$ 
      \If { not $done$ }
        \State $\epsilon_{last} = \epsilon_{min}$
        \State $\mlam_{kept} = \mlam_{kept} \cup \mlam_{min}$ {\hskip0.4in} (update kept symbols)
      \EndIf
    \EndWhile
    \vskip 10pt
    \State {\bf Output:} $\{ \mlam_{kept} \}$ (the Greedy Basis)
  \end{algorithmic}
\end{algorithm}
}
\par The very similar ``negative'' greedy algorithm removes model symbols until representation error increases beyond a specified threshold.
%
%
%
%
\subsection{Greedy Multivariate Polynomial Fitting}
The study of smooth scalar functions (e.g. $f(\vec{x})$) often centers about the Taylor series expansion.
In that instance, it is clear that any infinitely differentiable scalar function of many variables can be represented in terms of its derivatives by
\begin{align}
  \label{eq:mvt}
  f( \vec{x} + \vec{h} ) &= e^{ \vec{x} \cdot \vec{\nabla}' } \, f(\vec{x}\,') |_{\vec{x}\,'=\vec{h}}
  \\ \nonumber
  &\approx \sum_{k=0}^{K} \, \frac{1}{k!} \, (\vec{x} \cdot \vec{\nabla}')^{k} \, f(\vec{x}\,') |_{\vec{x}\,'=\vec{h}}
\end{align}
From the first to second line of \eqn{eq:mvt}, we have used the definition of the exponential function (i.e. its series expansion).
In the second line, the equality has been replaced by an approximation as we have limited the linear representation to $K+1$ terms.
\par This latter point is key to the perspective of \gmvp{}:
given training data thought to be related to a smooth multivariate function, it may, particularly on small scales, be well approximated by a truncated series expansion in an appropriate coordinate basis.
%
\hspace{1cm}
{\scriptsize
\begin{algorithm}[H]
  \caption{\gmvp{}, a degree tempered stepwise algorithm for multivariate polynomial modeling of scalar data.}
  \label{alg:gmvp}
  \begin{algorithmic}[1]
    \State {\bf Input:} $ \{ x, f, $max\_degree = 6$, tol \}$
    \vskip 10pt
    \State Define, $\mlam_{bulk}$, the bulk symbol space, to be the set of all multinomial combinations of basis vectors up to a predefined maximum order.
    \State Define $\mathcal{A}_{\gmvp{}}$ according to Alg. (\ref{alg:A_gmvp}).
    \State Given $max\_degree$, define, ${ D}$, a list of allowed polynomial degrees (e.g. $\{0,1,2,3,4,5,6\}$)
    \For {$d$ \textbf{in}  ${ D}$ }
      \State Define $\mlam_{bulk}^{(d)}$ as all symbols from $\mlam_{bulk}$ with degree less than or equal to current degree: $\mlam^{(d)}_{bulk}$
      \State Using $\mlam_{bulk}^{(d)}$, apply Alg. (\ref{alg:pgreedy}), \pgreedy{}, with $\mathcal{A}_{\gmvp{}}$ to get symbol subset, $\mlam^{(d)}_{opt}$ and estimator val, $\epsilon^{(d)}_{opt}$
      \If { $|\epsilon^{(d)}_{opt}-\epsilon^{(d-1)}_{opt}|<tol$ }
        \State \textbf{break}
      \EndIf
    \EndFor
    \vskip 10pt
    \State {\bf Output:} $\mlam^{(d)}_{opt}$
  \end{algorithmic}
\end{algorithm}
}
\par \eqn{eq:mvt}'s truncated expansion happens to be a polynomial in at most $N$ variables.
In this setting, the uncertainty of which and how many basis terms to include makes this a problem ripe for the application of linear modeling driven by a greedy process, namely \eqn{eq:linsol1} and Alg. (\ref{alg:pgreedy}) .
%
%
\begin{figure*}[htb]
  \begin{tabular}{ll}
    {\includegraphics[width=0.49\textwidth]{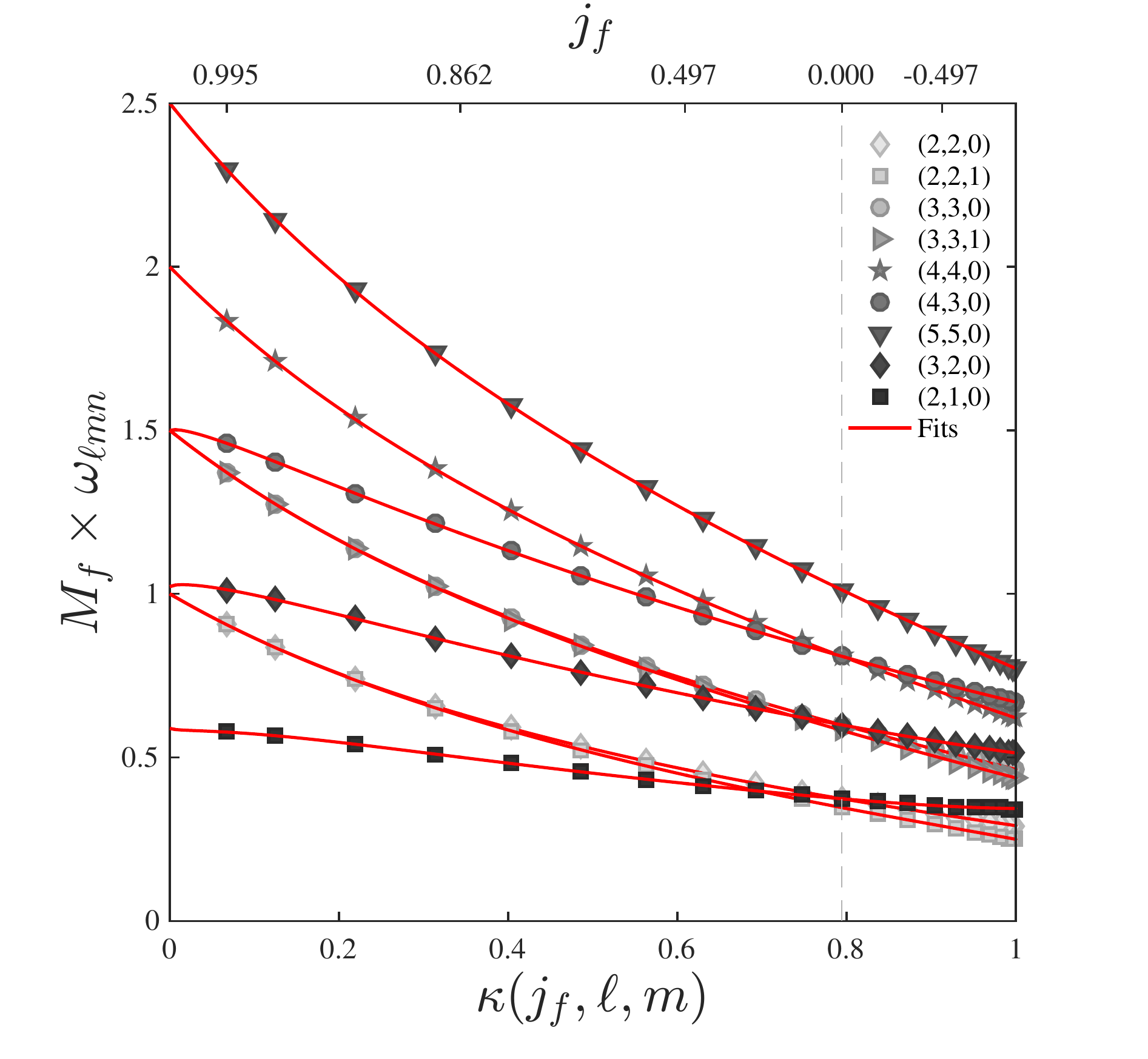}} & {\includegraphics[width=0.49\textwidth]{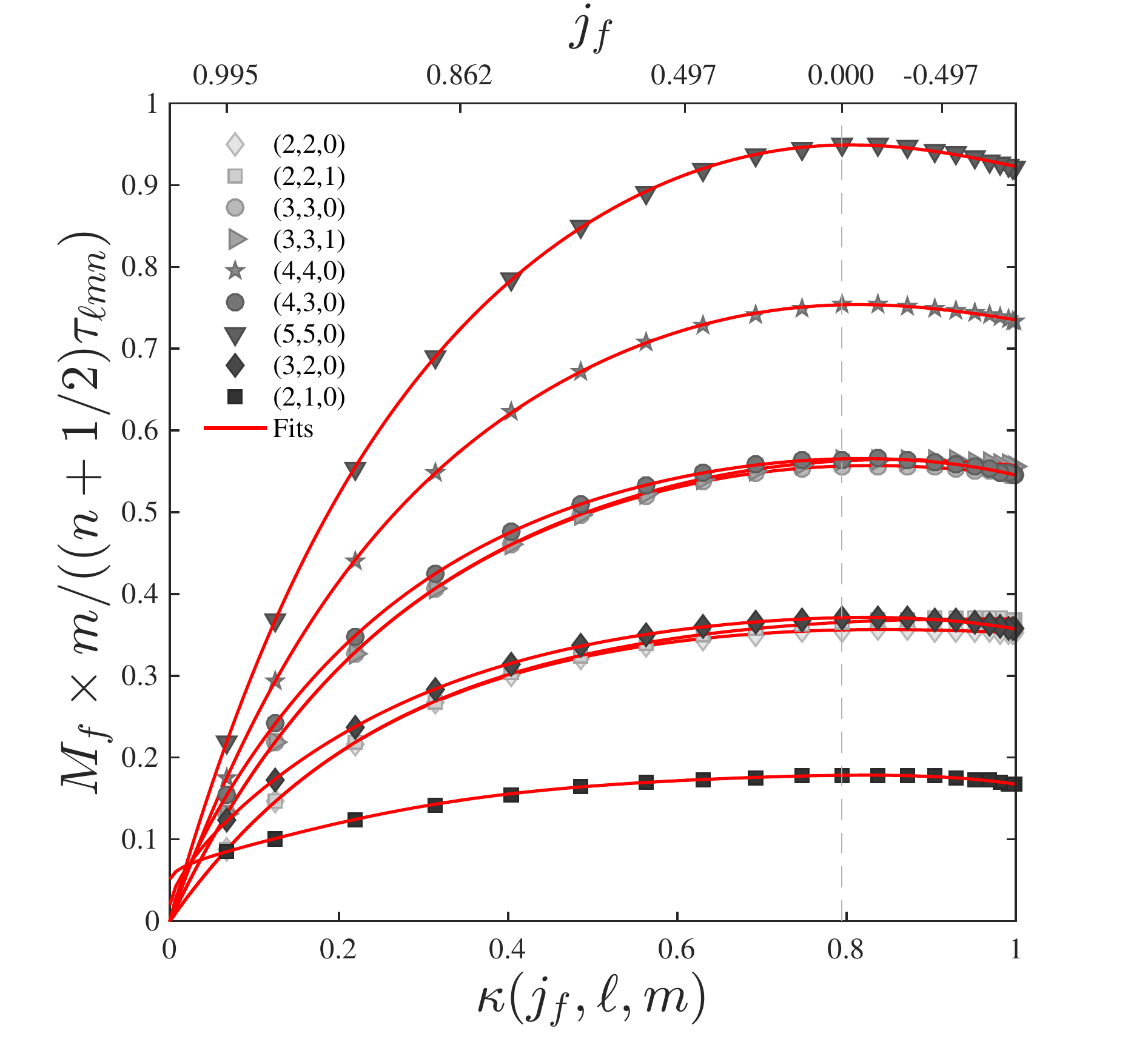}}
    \\
    \hspace{0.35cm}\includegraphics[width=0.43\textwidth]{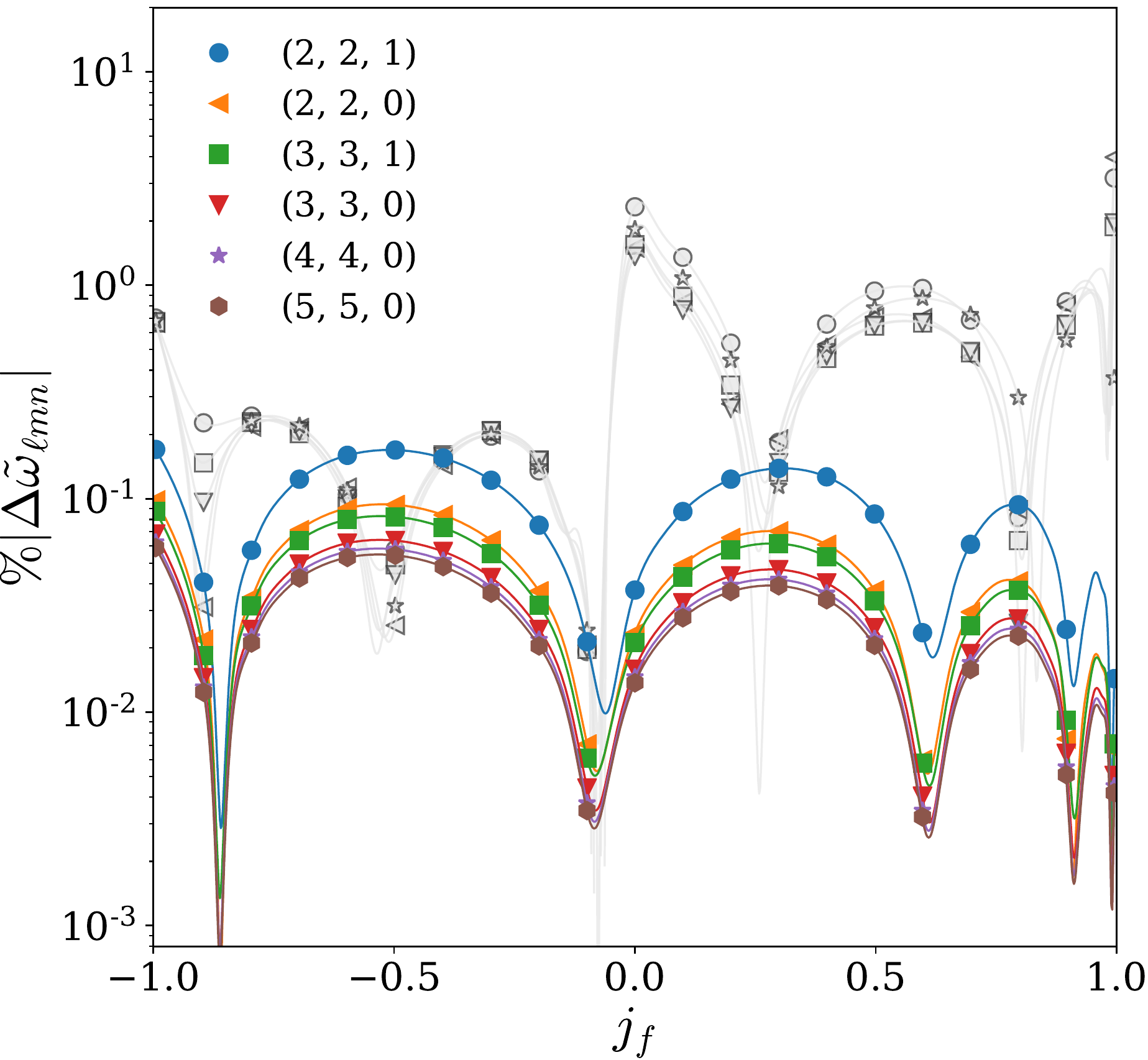} & \hspace{0.3cm}\includegraphics[width=0.43\textwidth]{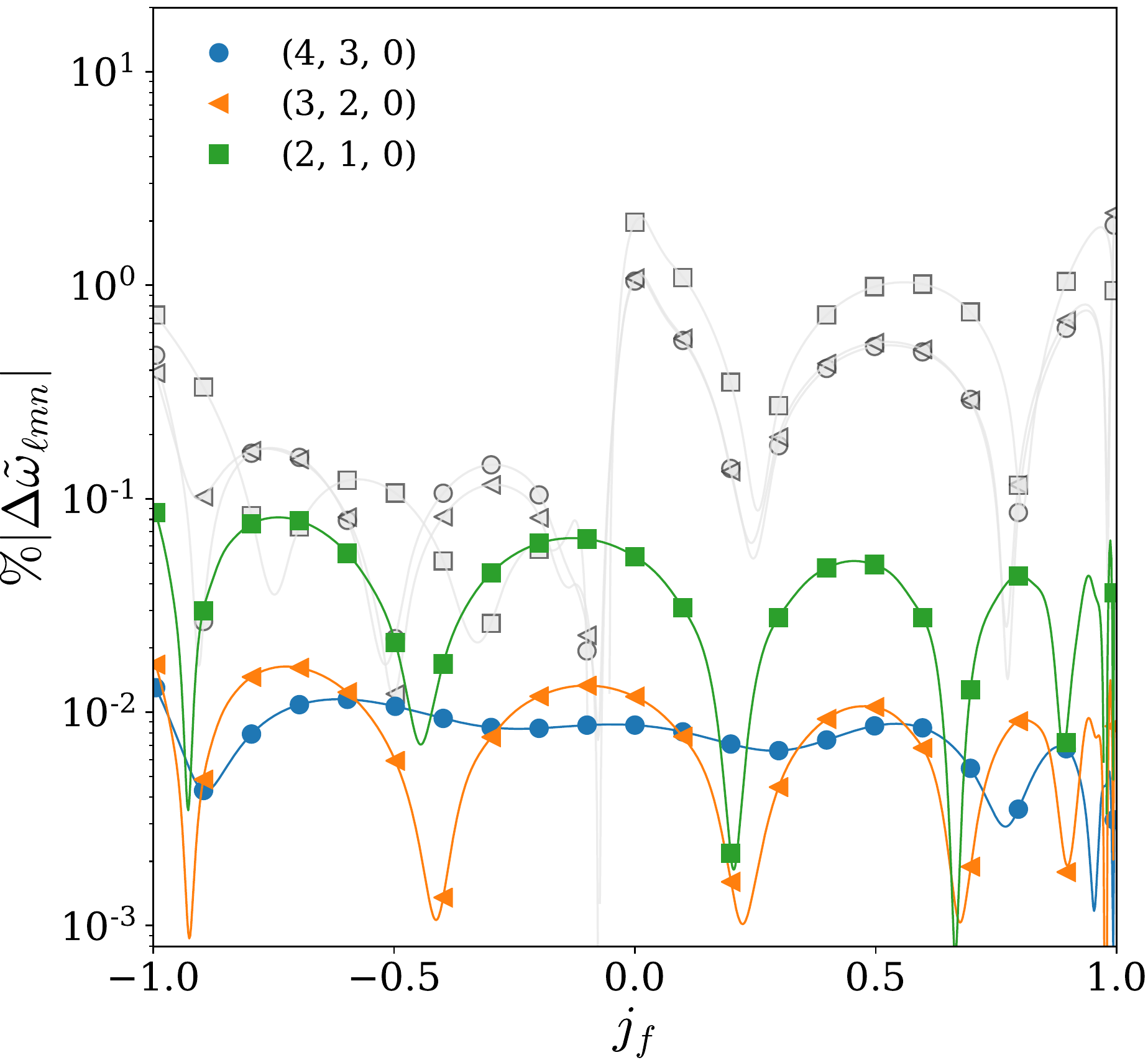}
  \end{tabular}
	\caption{ Fits of dimensionless \qnm{} central frequencies (solid lines) along with select numerical values (grey markers) computed using Leaver's method \cite{Leaver85}.
  Before the application of $\kappa(\j{})$, points are spaced between -\CwFitCalibrationRegion and \CwFitCalibrationRegion according to \CwFitCalibrationRegion times the $\sin$ of a fiducial angle which is uniformly spaced between $-\pi/2$ and $\pi/2$. Values of $\j{}$ are shown in the upper axis for $\kappa$ at $l=m$.
  The grey dashed line marks the value of $\kappa$ where $\j{}=0$. Fits of dimensionless \qnm{} decay rates (solid lines) along with select numerical values (grey markers) computed using Leaver's method \cite{Leaver85}. (Bottom) Percentage absolute residual errors for fits of dimensionless \qnm{} frequencies,, $\cw\lmn$ along with select numerical values (colored markers) computed using Leaver's method \cite{Leaver85}. For comparison, residuals using the models in Ref.~\cite{Berti:2005ys} are shown in gray. }
  \label{fig:qnm}
\end{figure*}
\par Here, the basis symbols required by Alg. (\ref{alg:pgreedy}) are the multinomial terms in \eqn{eq:mvt}.
Each term is an element of the tensor-product of the power-sets of each model dimension. That is $\mlam_{bulk} = \{x_0,x_1 .. x_N, ... x_0^1, x_0^2, x_0^3, ...  , (x_0^K x_1^K ... x_N^K) \}$.
Note that in practice it may be useful to encode elements of $\mlam_{bulk}$ with strings representing their constituents (e.g. $x_0 x_0 x_1 x_2 x_2 x_4$ could be represented by the string ``001224''). This provides a way of bijectively mapping between symbols and numerical basis vectors.
\par The action, $\mathcal{A}(\mlam_{trial})$, required by Alg. (\ref{alg:pgreedy}) encompasses the evaluation of \eqn{eq:linsol1} to solve for the basis coefficients, $\mu_k$, and the calculation of the modeling error.
An explicit sketch of this is given by Alg. (\ref{alg:A_gmvp}).
\par The combination of these two ideas alone results in an algorithm prone to a deficit of stepwise methods: the algorithm may confuse correlated basis vectors (e.g. $x^2$ may be confused with $x^4$).
To counter this, we may incrementally increase, or \textit{temper}, the maximum allowed multinomial degree.
For example, when iterating through allowed degrees, if the current maximum degree is 3, then degree 4 terms, such as $x_0x_1x_3^2$, will not be considered within the space of model symbols.
The degree tempering process halts when increasing the maximum allowed degree has no significant effect on model representation error.
\par The combination of degree tempering with the greedy approach results in the \gmvp{} algorithm as presented in Alg. (\ref{alg:gmvp}).
\subsection{Greedy Multivariate Rational Fitting}
\par Despite the apparent universality of \eqn{eq:mvt}, there are many cases where $K$ must be orders of magnitude greater than 1 in order for $\vec{f}$ to be accurately represented by a low order multinomial.
Worst, in cases where the underlying dataset is best described by a rational function, no polynomial of the same family describing the numerator and denominator will yield satisfactory results in low order (e.g. the well known ``Runge's phenomenon'' \cite{10.2307/2323093}).
In general, the optimal polynomial basis may not be clear, and so a more general set of ansatzes may be of use.
\par Of the simplest of such ansatzes are rational functions of the form
\def\muf{\bar{\mu}}
\def\sif{\bar{\sigma}_f}
\begin{align}
  \label{eq:rat1}
  f(\vec{x}) = \muf + \sif \; \frac{ \sum_{r=0}^{R} a_r \, \phi_{r}(\vec{x}) }{ 1 - \sum_{v=1}^{V} b_v \, \phi_{v}(\vec{x}) } \;,
\end{align}
where $\muf$ is the additive mean of $f(\vec{x})$, and $\sif$ is the standard deviation of $f(\vec{x})$, and $\phi_k$ are the multinomials basis functions considered in the previous section.
Note that, in \eqn{eq:rat1}, the sum over $v$ does not include the constant term associated with $\phi_0$.
\par While it is tempting to embrace \eqn{eq:rat1}'s $f({\vec{x}})$ as a nonlinear function and so resort to nonlinear modeling methods, a reformulation reveals an underlying linear structure \cite{Press:1992:NRC:148286}.
Namely, if we let
\begin{align}
  g = (f - \muf)/\sif
\end{align}
then algebraic manipulation of \eqn{eq:rat1} allows
\begin{align}
  \label{eq:rat2}
  g = \sum_{r=0}^{R} a_r \, \phi_{r}(\vec{x}) + g\, \sum_{v=1}^{V} b_v \, \phi_{v}(\vec{x}) \; .
\end{align}
We are free to relabel the indices such that \eqn{eq:rat2} is manifestly linear in a single index.
At this stage, we will also explicitly consider the $j^\mathrm{th}$ samples of the domain, and so refer to (e.g.) $\vec{x}$ as $\vec{x}_j$.
\begin{figure*}[htb]
  \includegraphics[width=\textwidth]{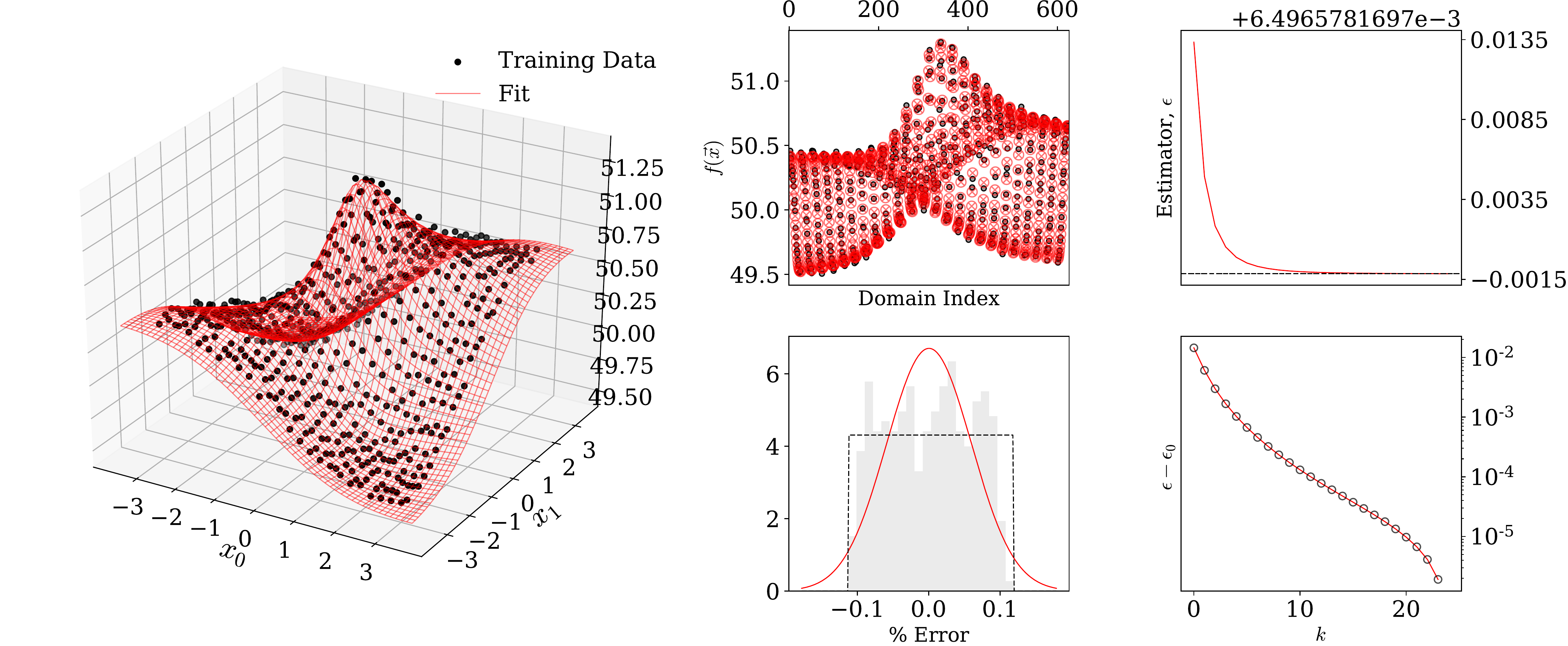}
	\caption{ Standard summary plot for Greedy Multivariate Rational fitting algorithm (\gmvr) as implemented in \cite{lionel_london_2018_1402516}. (\textit{left}) 3D plot of training data (black dots) and final fit (red mesh). (\textit{center top}) Same as left most panel, but in index space. (\textit{center bottom}) {Percent residual error with respect to validation data (grey blocks) along with uniform (black) and gaussian (red) fits to error. The validation data were generated in the same manner as the training data.} (\textit{top right}) convergence of the $L^2$ norm during iterative refinement. (\textit{bottom right}) Same as top right, but on log scale, where $\epsilon_0$ is the value of $\epsilon$ at the final $k^{\mathrm{th}}$ iteration of refinement. }
  \label{fig:gmvrtoy}
\end{figure*}
These adjustments of perspective result in
\begin{align}
  \label{eq:rat3}
  g_j = \sum_{k=0}^{R+V} \, z_{k} \, \psi_k(\vec{x}_j) \;,
\end{align}
where
\begin{align}
  z_k = \left\{ \begin{array}{cc}
        a_k,      & \text{for }0\leq k\leq R\\
        b_k, & \quad \;\; \;\text{for }R+1\leq k\leq R+V
      \end{array} \right\}
\end{align}
and
\begin{align}
  \psi_k(\vec{x}_j) = \left\{\begin{array}{cc}
        \phi_k(\vec{x}_j),      & \text{for }0\leq k\leq R\\
        \phi_k(\vec{x}_j)\, g_j , & \quad \;\; \;\text{for }R+1\leq k\leq R+V
        \end{array} \right\}
\end{align}
\par Recalling \eqns{eq:linsol1}{pinv1}, it follows that the coefficients of interest ($a_k$ and $b_k$), may be estimated according to
\begin{align}
  \label{eq:pinv2}
  \vec{\alpha} = \hat{P} \; \vec{g} \;.
\end{align}
where, $\hat{P}$ is the pseudo-inverse of the matrix whose elements are $\psi_k(\vec{x}_j)$, $\vec\alpha = ( z_0, z_1, ... z_{R+V-1},z_{R+V} )$, and $\vec{g} = ( g_0, g_1, ... g_{R+V-1},g_{R+V} )$.
\par However, we note that $\hat{P}$ depends nontrivially on $g$, and is therefore susceptible to noise in the training data.
Let us briefly consider the effect of zero-mean noise on $g$, e.g. $g \rightarrow g+n$.
In this, it may be that shown that $n$ may be entirely relegated to $\hat{P}$.
It is in this sense that \eqn{eq:pinv2} is insufficient to generally solve for $\vec{\alpha}$, as $\hat{P}$ may be adversely affected by noise.
\par The key to robustly solving for $\vec{\alpha}$ lies in iterative refinement \cite{Press:1992:NRC:148286}.
Specifically, we note that \eqn{eq:rat2} may be modified to iteratively minimize the impact of numerical noise on $\hat{P}$.
That is, to reduce the impact of noise on $\hat{P}$, we are free to calculate it using model evaluations of $g$ rather than the original (noisy) training data.
If we define $g^{(0)}=g$ (i.e. $g$ is the training data), with $\hat{P} = \hat{P}( \vec{g}^{\,(n)} )$, then \eqn{eq:pinv2} generalizes to
\begin{align}
  \label{eq:pinv3}
  \vec{\alpha}^{(n+1)} = \hat{P}(\vec{g}^{\,(n)}) \, \vec{g}^{(0)}\;.
\end{align}
In practice, one solves \eqn{eq:pinv3} for $\vec{\alpha}^{(n+1)}$, and then uses the related $a_r^{(n+1)}$ and $b_v^{(n+1)}$ to calculate $g^{(n+1)}$ via
\begin{align}
  \label{eq:rat4}
  g^{(n+1)} =  \frac{ \sum_{r=0}^{R} a^{(n+1)}_r \, \phi_{r}(\vec{x}) }{ 1\;-\;  \sum_{v=1}^{V} b^{(n+1)}_v \, \phi_{v}(\vec{x}) } \; .
\end{align}
Subsequently, $g^{(n+1)}$ is then fed back into \eqn{eq:pinv3} for further refinement.
The refinement process is to terminate when a measure of model error (e.g. the $L^2$ norm $||g^{(0)}-g^{(n)}||$) passes a predetermined threshold.
%
%
%
{\scriptsize
\begin{algorithm}[H]
  \caption{$\mathcal{A}_{\gmvp}$, the action for \gmvp. Model calculation given basis symbols, and output of model error estimate.}
  \label{alg:A_gmvp}
  \begin{algorithmic}[1]
    \State {\bf Input:} $\mlam_{trial}$
    \vskip 10pt
    \State Calculate $\mu_k$ via \eqn{eq:linsol1}.
    \State Calculate the model representation error, e.g.: $\epsilon = ||\hatU \vecmu - \vec{f}||/||\vec{f}||$, where $||a||$ is the $L^2$ norm of $a$.
    \vskip 10pt
    \State {\bf Output:} $\epsilon$
  \end{algorithmic}
\end{algorithm}
}
\par Much as in the case of multivariate polynomial fitting, we are left with an unknown number and content of basis symbols.
In principle, the existence of $a_r$ and $b_v$ makes the problem more complicated, as one might imagine optimizing over each symbol space independently.
To broach this complications, we again use a greedy algorithm with degree tempering.
However, rather than independent greedy optimizations for the numerator and denominator bases symbols, \eqn{eq:rat3} suggests that the appropriate labeling of symbols (e.g. ``numerator'' or ``denominator'') may yield an effective flattening of the supposed 2D symbol selection problem.
Put another way, rather than two simultaneous greedy optimizations over $R+1$ and $V$ symbols (with $(R+1)V$ iterations), a single greedy process over $V+R+1$ symbols is performed, where each symbol is additionally labeled as being in the numerator or denominator.
\par With these conceptual tools in hand, we may proceed to constructing \gmvr{} by first defining its action, $\mathcal{A}_{\gmvr}$.
This is done in Alg. (\ref{alg:A_gmvr}).
\par The combination of \eqn{eq:pinv3} and \eqn{eq:rat4}, along with \pgreedy{} and degree tempering, results in the \gmvr{} algorithm as presented in Alg. (\ref{alg:gmvr}).
Both \gmvp{} and \gmvr{} are publicly available on Github through the \texttt{positive} repository (Ref. \cite{lionel_london_2018_1402516}), and may be imported in python via {\small{\texttt{positive.learning.gmvpfit}}} and {\small{\texttt{positive.learning.gmvrfit}}}.
\section{Results}
\label{results}
\par We briefly review the application of \gmvr{} to a toy problem wherein a scalar rational function of two variables is treated.
%
%
%
We then present two applications to \gw{s}.
First we apply \gmvp{} to the modeling of complex valued Kerr \qnm{} frequencies.
Second, we apply \gmvr{} to the modeling of spin -2 spherical-spheroidal harmonic mixing coefficients (\ceqn{sigma}).
While only 1D and 2D domains are treated here, we note that Ref. \cite{London:2018gaq} has used a version of \gmvp{} to model the \qnm{} excitation amplitudes in a 4D parameter space.
\subsection{\gmvr{} Toy Problem}
Here, our goal is to very briefly overview the functionality of the \gmvr{} algorithm as implemented in Ref. \cite{lionel_london_2018_1402516}.
While it is possible to investigate the output of \gmvr{} with varying hyper-parameters (such as the tolerance input to Alg. \ref{alg:gmvr}), we will focus only on a simple usage case.
Similarly, we note that \gmvr{} as implemented in Ref. \cite{lionel_london_2018_1402516} involves a negative greedy phase to counter over-modeling in cases where the aforementioned $tol$ is too low.
For relevance of presentation to physics examples in subsequent sections, we will restrict ourselves to a case where numerical noise is low, and the negative greedy step does not alter the output of Alg. \ref{alg:gmvr}.
\par Let us now consider the application of \gmvr{} to a fiducial scalar function of the form
\begin{align}
  f(x_0,x_1) = \mu + \sigma \, \left( \, \frac{a_0 +a_1  x_0 + a_2 x_1 + a_3 x_0 x_1}{ 1 + b_1 \, x_0^2 +  b_2 x_1^2 } \, \right) + 0.05 \, n \; ,
\end{align}
where $n$ is a uniform random variable on $[-1,1]$.
Towards easily identifying test values for $a_j$ and $b_k$ with those recovered, it is more straightforward to distribute $\sigma$ to the denominator, yielding
\begin{align}
  \label{eq:gmvrtoy1}
  f(x_0,x_1) = \mu + \, \frac{a_0 +a_1  x_0 + a_2 x_1 + a_3 x_0 x_1}{ 1/\sigma + (b_1/\sigma) \, x_0^2 +  (b_2/\sigma) x_1^2 } + 0.05 \, n \; .
\end{align}
Under this perspective we will consider test data generated with the parameters listed in \tbl{tb:gmvrtoy}'s left two panels.
\par To generate the test data, \eqn{eq:gmvrtoy1} is evaluated with 25 points along $x_0$ and $x_1$ (with $25^2$ total points), where each is between -3 and 3. Though not a requirement of \gmvr{}, for simplicity of presentation, domain points are equally spaced.
\par \fig{fig:gmvrtoy} shows the application of \gmvr{} to this fiducial dataset.
\fig{fig:gmvrtoy}'s central bottom panel displays the distribution of percentage residuals with respect to validation data generated in the same manner as training data.
A gaussian fit to the fractional residuals is displayed for comparison.
In particular, despite the uniform nature of the underlying noise distribution, a biased fit will often have residuals that are approximately gaussian.
We see that this is not the case here, and that the uniformly random noise distribution is approximately recovered.
Moreover, when considering many noise realizations to generate validation data, we find that sample noise and residuals have an average correlation of $99.46\%$.
\par \fig{fig:gmvrtoy}'s right top and bottom panels show the convergence of Alg. (\ref{alg:A_gmvr})'s iterative refinement stage (i.e. its while-loop).
Here it is demonstrated that \gmvr{} converges in a way that is approximately exponential, owing to the underlying analytic nature of the training data.
\tbl{tb:gmvrtoy} demonstrates \gmvr{}'s accurate recovery of the underlying model parameters.
We note that \gmvr{}'s initial output contains terms in the numerator which correspond to the addition of a constant to the overall model, thus correcting for the difference between the offset parameter, $\mu$, and the true, but arbitrary, mean of the dataset.
\tbl{tb:gmvrtoy} presents recovered model parameters after this effect has been accounted for with simple algebraic manipulation.
%
%
%
\begin{table}
  \caption{Summary of recovered model parameters for \gmvr{} toy problem.}
  \label{tb:gmvrtoy}
  \begin{tabular}{|c|c|c|c|}
    \hline
    \hline
    Parameter & Training Value & Modeled Value & Difference \\ \hline \hline
    $\mu$ & 50.0 & 49.9915 & 0.0171 \% \\ \hline
    $a_0$ & 1.1 & 1.1374 & 3.4002 \% \\ \hline
    $a_1$ & 0.2 & 0.2000 & 0.0000 \% \\ \hline
    $a_2$ & 0.5 & 0.5068 & 1.36784 \% \\ \hline
    $a_3$ & 1.0 & 1.0063 & 0.6300 \% \\ \hline
    $1/\sigma$ & 0.9 & 0.9375 & 4.1612 \% \\ \hline
    $b_1/\sigma$ & 1.0 & 0.9941 & 0.5906 \% \\ \hline
    $b_2/\sigma$ & 1.0 & 1.0000 & 0.0000 \% \\ \hline
    \hline
  \end{tabular}

\end{table}
\par In this rudimentary example case, \gmvr{} correctly recovers the functional form of the input data, and accurately recovers the correct values of model parameters.
But, in general, \gmvr{} and related techniques, having no knowledge of the underlying noise distribution, will attempt to model minor correlations and offsets within the training data's noise.
However, we have demonstrated the utility of \gmvr{} in a relatively ideal usage case where the underlying function is rational, and the training data is only weakly contaminated with noise.
\par In the following sections, we consider realistic, but similarly ideal cases, where the functional form of the sample data is not known to be explicitly polynomial or rational, but the amount of noise within the training data is negligible.
\begin{figure}[htb]
  \includegraphics[width=0.45\textwidth]{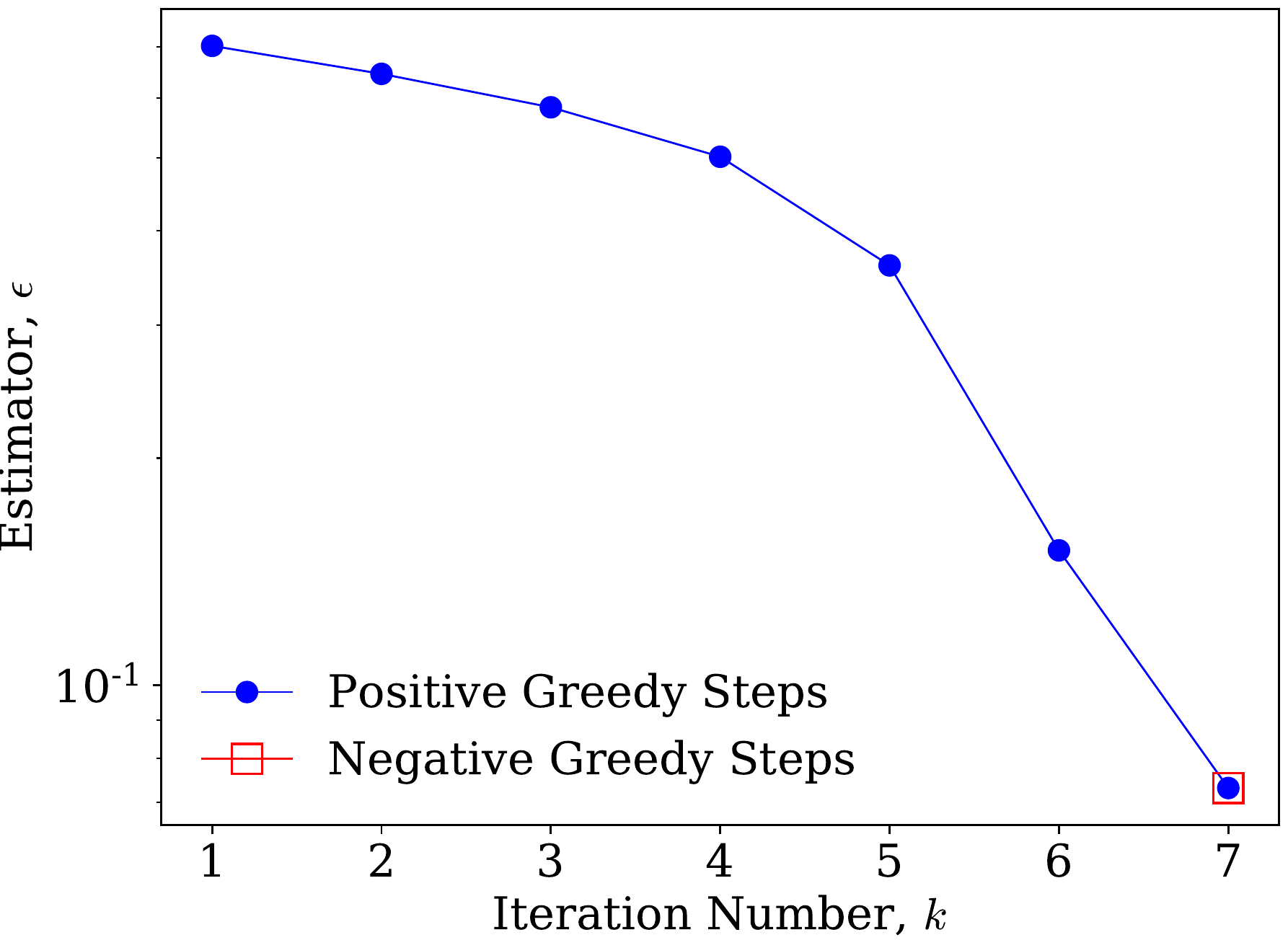}
	\caption{ Convergence of greedy process for \gmvr{} toy problem. }
\end{figure}
%
%
%
%
\subsection{Modeling \qnm{} frequencies with \gmvp{}}
%
%
%
In seeking to apply \gmvp{} to select \qnm{} frequencies, we wish to account for the known extremal Kerr behavior of some modes.
Namely, we will impose a zero-damping constraint: some frequencies are real as $\jf \rightarrow 1$ \cite{Zimmerman:2015trm}.
We also wish to impose a domain transformation, $\kappa(\jf,\ell,m)$, such that $0 \leq \kappa \leq 1$ and the individual \qnm{} frequencies are made approx. polynomial in $\kappa$.
\par For the domain transformation, inspection of \qnm{} frequencies with $\ell \leq 5$ suggest that
\begin{align}
  \label{eq:kappa}
  \kappa(\jf,\ell,m) = \left( \log_3( 2 - \jf ) \right)^{1/(2+\ell-|m|)}
\end{align}
appropriately linearizes the sharp behavior of each frequency near $\jf=1$ while also mapping $-1 \le \jf \le 1$ onto $0 \le \kappa \le 1$.
\par Towards the zero-damping constraint, when considering a \qnm{} frequency $\cw_{\lmn}$, zero-damping at $\jf=1$ implies that $\cw_{\lmn}(\kappa \approx 0)-m/2 \propto \kappa$, where $m/2$ is the well known limiting value for each \qnm{} frequencies real part as $\jf\rightarrow 1$.
This implies that
\begin{align}
  \label{eq:zd}
  \cw_{\lmn} = m/2 + \kappa \sum_{j=1}^{J} c_j \kappa^k \; .
\end{align}
In the case of the non-zero damped \qnm{s}, (e.g. $(\ell,m)=(3,2)$), a more general polynomial form may be adopted, namely,
\begin{align}
  \label{eq:nzd}
  \cw_{\lmn} = \sum_{j=0}^{J} c_j \kappa^k \; .
\end{align}
\par The polynomial content of \eqn{eq:zd} and \eqn{eq:nzd} is determined by \gmvp{}.
\eqns{eq:cw_fit_1}{eq:cw_fit_9} display the resulting polynomial models.
In particular, the domain map allows most \qnm{} frequencies to be well modeled by 4th order polynomials which include all lower degree terms; concurrently, the real and imaginary parts of each $\cw\lmn$ are modeled simultaneously.
\par \fig{fig:qnm} displays select training points, as well as model fits for $\cw\lmn$'s real and imaginary parts.
For the top right and top left panels, the simple polynomial behavior of each curve is a result of the displayed linear domain in $\kappa(\jf,\ell,m)$.
In the top left panel, we have scaled $1/\tau\lmn$ by factors for $m/(n+1/2)$ to place the \qnm{s} with $n=0$ and $n=1$ at approximately the same scale.
%
\begin{widetext}
	\input{cwfit_eqns}
\end{widetext}
\par \fig{fig:qnm}'s top left and right panels' upper axes demonstrate the effect of mapping $\jf$ onto $\kappa$.
In particular, it is shown that the two branches (namely $\jf>0$ and $\jf<0$) naturally form a single family of solutions when accounting for the sign of the \bh{}'s oriented spin \cite{Husa:2007hp}.
Concurrently, the use of $\kappa$ as a domain variable has the desirable effect of making each $\cw\lmn$ and $\tau\lmn$ approximately polynomial.
We note that, in the asymptotic vicinity of $\jf=1$, the \qnm{} frequencies and decay times are known to have solutions that are asymptoticly degenerate \cite{Zimmerman:2015rua}.
{\scriptsize
\begin{algorithm}[H]
  \caption{$\mathcal{A}_{\gmvr}$, the action for \gmvr. Model calculation given basis symbols, and output of model error estimate.}
  \label{alg:A_gmvr}
  \begin{algorithmic}[1]
    \State {\bf Input:} \{ $\mlam_{trial}$, $tol=10^{-3}$ \}
    \vskip 10pt
    \State Calculate $\alpha^{(1)}_k$ via \eqn{eq:pinv3}. Implicitly, $n=1$.
    \State Calculate the current model prediction $g^{(1)}$ via \eqn{eq:rat4}.
    \State Calculate the model representation error, e.g.: $\epsilon^{(1)}  = ||var(g^{(n)} - g^{(0)})||/||g^{(0)}||$, where $var$ is the variance.
    \State $done = False$
    \While { not $done$ }
      \State Calculate $\alpha^{n+1}_{k}$  via \eqn{eq:pinv3}.
      \State Calculate the current model prediction $g^{(n)}$ via \eqn{eq:rat4}.
      \State Calculate the model representation error, e.g. $\epsilon^{(n)} = ||g^{(n)} - g^{(0)}||/||g^{(0)}||$.
      \State $done = |\epsilon^{(n-1)}-\epsilon^{(n)}| < tol$
      \State (Implicitly, $n = n+1$)
    \EndWhile
    \vskip 10pt
    \State {\bf Output:} $\epsilon^{(n)}$
  \end{algorithmic}
\end{algorithm}
}
%
In allowing \eqns{eq:cw_fit_1}{eq:cw_fit_9} to extrapolate to $\jf=1$, we do not explicitly account for this additional effect.
\par \fig{fig:qnm}'s bottom two panels show absolute fractional residual errors of the complex frequency, $\cw\lmn = \omega\lmn + i/\tau\lmn$.
Although each model's fractional error is within 1\% of the perturbation theory result, each is dominated by systematic error due to the choice fitting ansatz.
For comparison, the same residual errors are shown in gray for the model presented in Ref.~\cite{Berti:2005ys};
here, the sharp feature near $\jf=0$ results from their modeling counter and co-rotating \qnm{} as two different curves.
\par Together, \eqns{eq:cw_fit_1}{eq:cw_fit_9} along with \fig{fig:qnm} present precise and accurate fits for the real and imaginary parts of \qnm{} frequencies for gravitational perturbations of Kerr \qnm{s}.
A \texttt{Python} implementation of \eqns{eq:cw_fit_1}{eq:cw_fit_9} is available in Ref.~\cite{lionel_london_2018_1402516} via \texttt{positive.physics.cw181003550}.
\subsection{Modeling spherical-spheroidal inner-products with \gmvr{}}
Here we apply \gmvr{} to the spherical-spheroidal mixing coefficients, $\sigma_{\LMlmn}$.
As in the case of the \qnm{} frequencies, we use the domain transformation defined by \eqn{eq:kappa} to simplify the functional form of each $\sigma_{\LMlmn}$.
\par While it is possible to enforce extremal Kerr and Schwarzschild limiting conditions for $\sigma_{\LMlmn}$, we find it effective to first use \gmvr{} to determine a functional form that works for {individual} $\sigma_{\LMlmn}$, and then from these ansatz develop a single ansatz for all $\sigma_{\LMlmn}$.
\eqns{eq:ys22220}{eq:ys55550} present the resulting model equations.
A \texttt{Python} implementation of \eqns{eq:ys22220}{eq:ys55550} is available in Ref.~\cite{lionel_london_2018_1402516} via \texttt{positive.physics.ysprod181003550}.
\par
\begin{figure*}[htb]
  \begin{tabular}{ll}
    \hspace{-0.21cm}\includegraphics[width=0.48\textwidth]{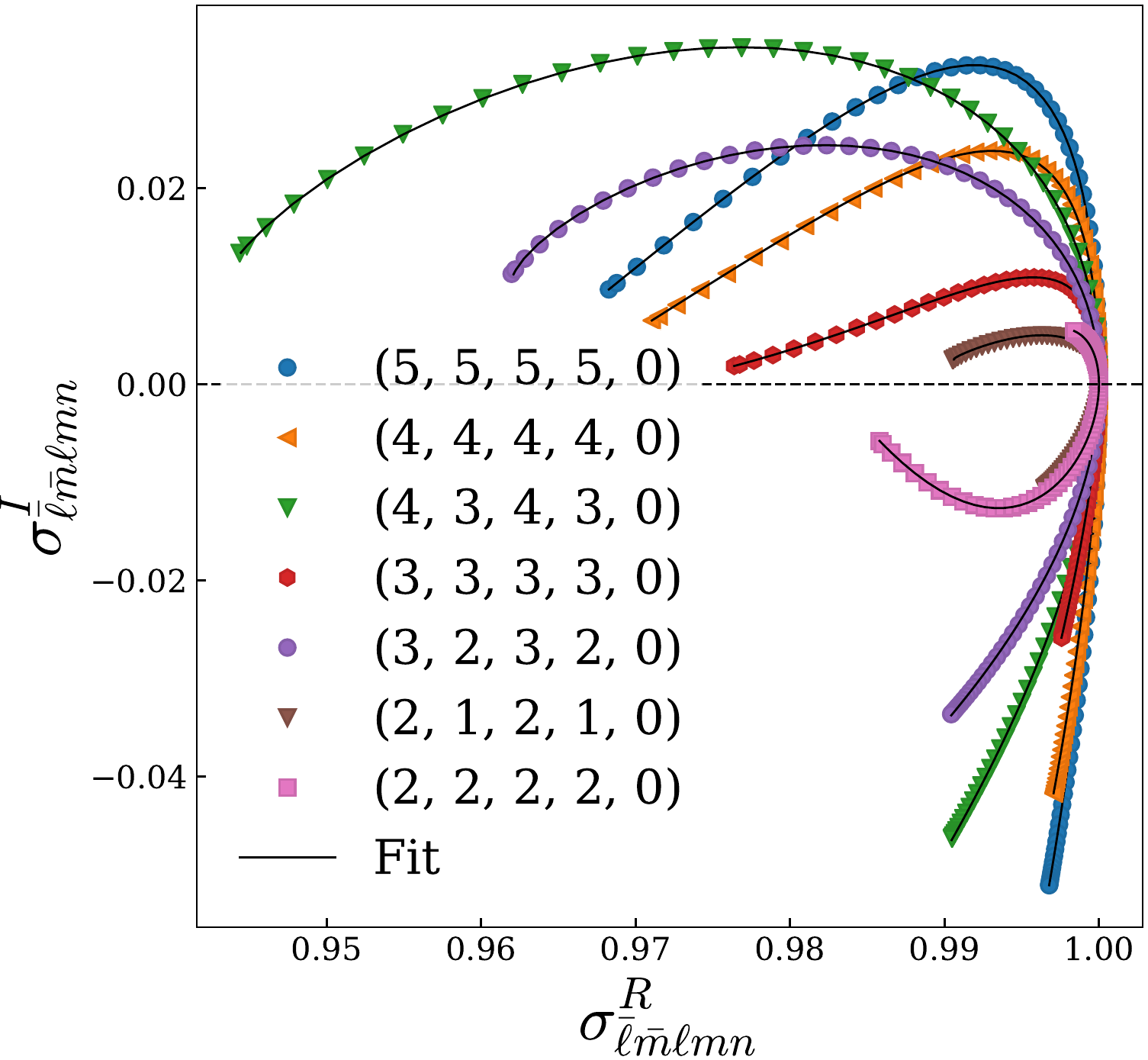} & \includegraphics[width=0.48\textwidth]{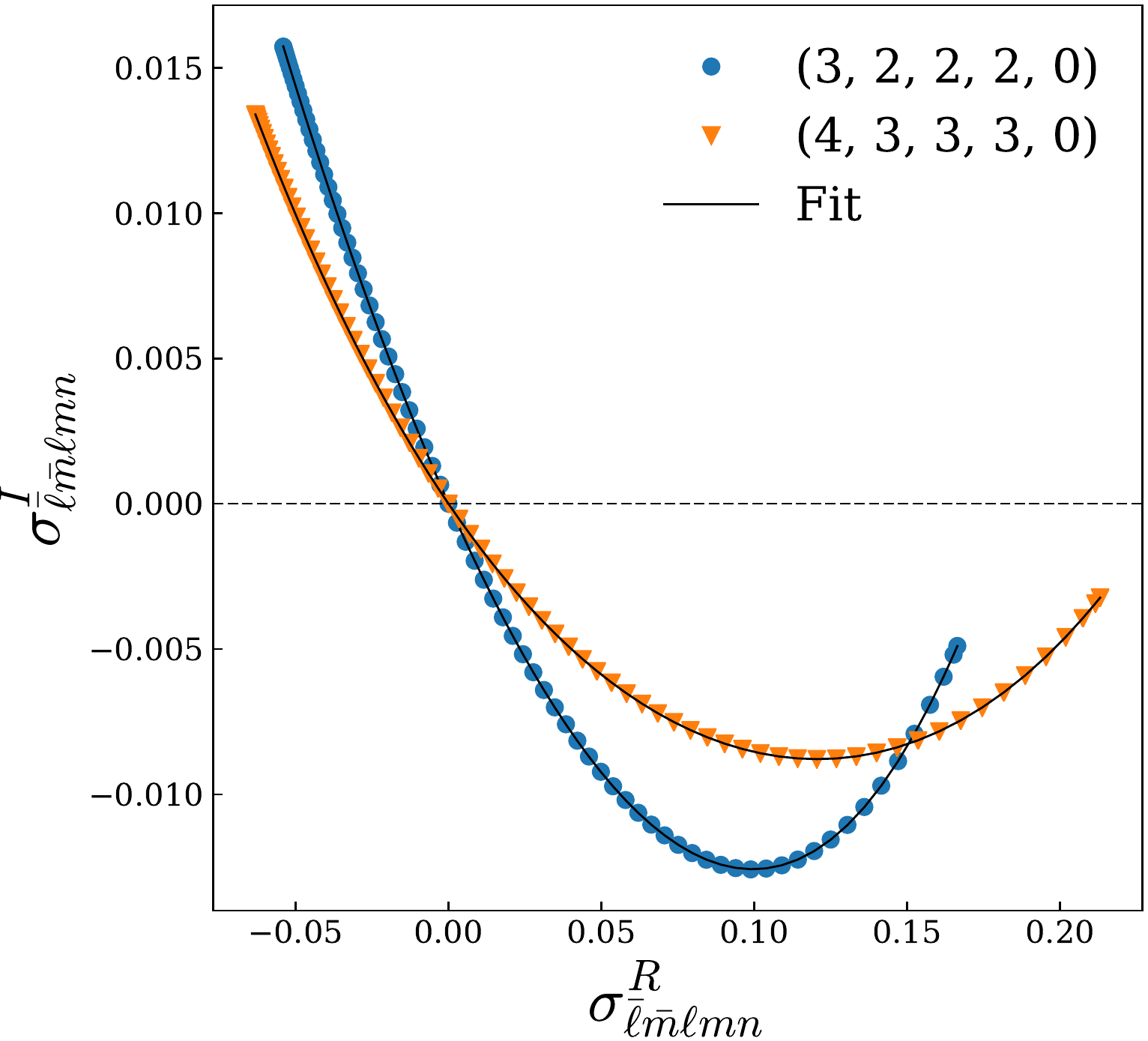} \\
    \includegraphics[width=0.48\textwidth]{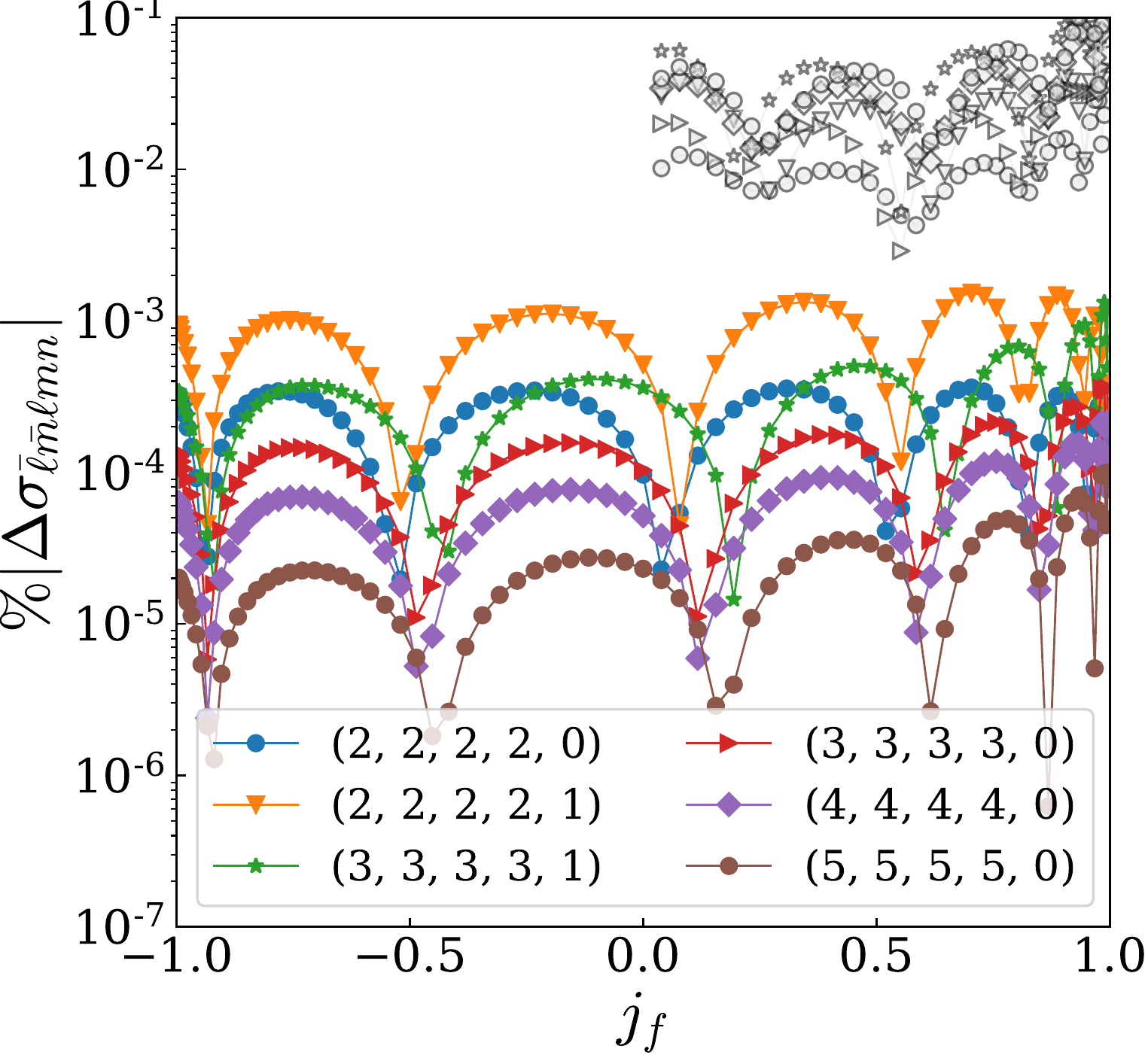} & \hspace{0.3cm}\includegraphics[width=0.48\textwidth]{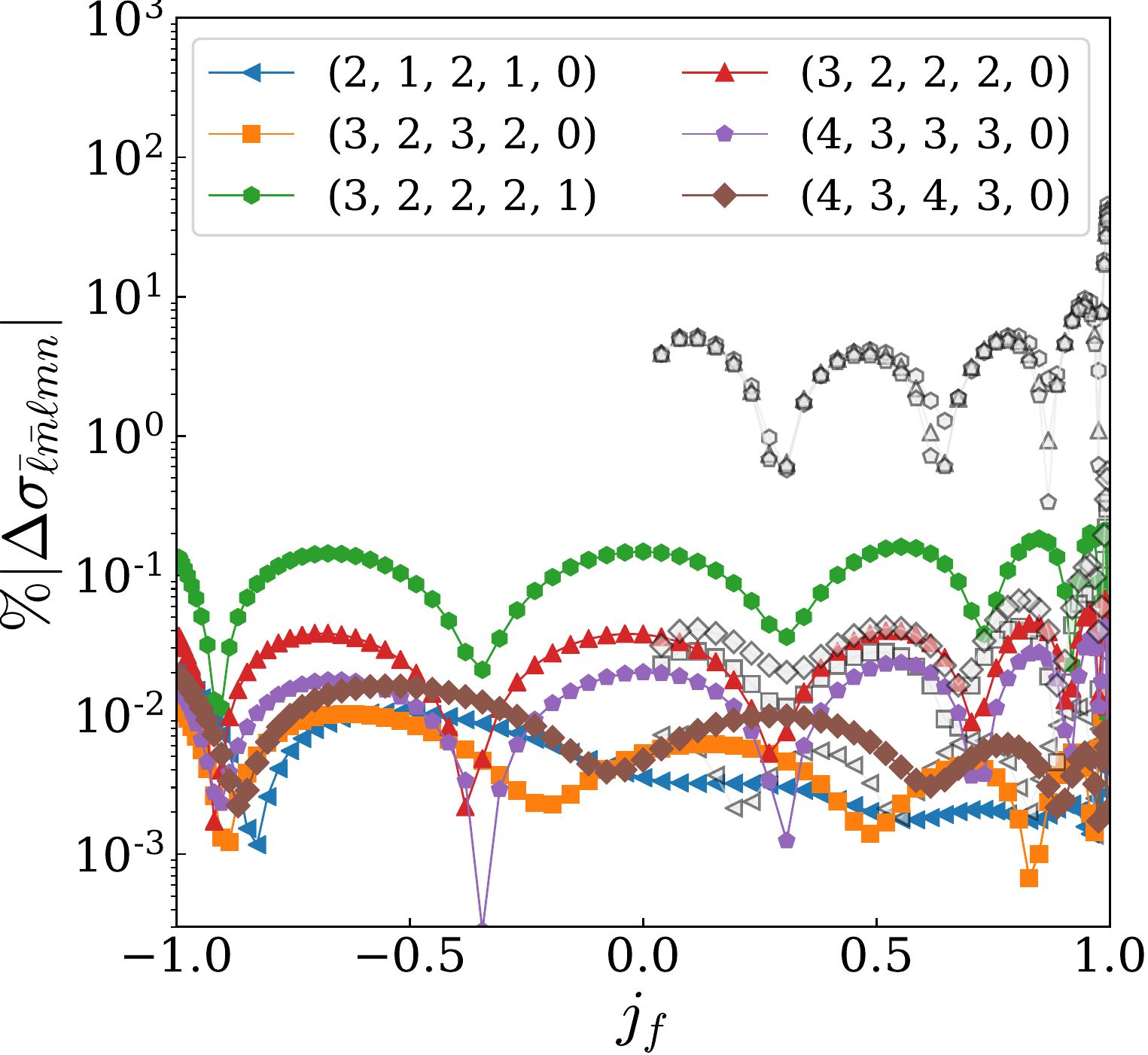}
  \end{tabular}
	\caption{ Spherical-spheroidal harmonic mixing coefficients and percentage residual errors for spherical-spheroidal harmonic mixing coefficient fits. (\textit{Top Left}) Mixing coefficients for cases where $(\bar{\ell},\bar{m})=(\ell,m)$. The dashed horizontal intersects each curve at the Schwarzschild limit where $\jf=0$. (\textit{Top Right}) Mixing coefficients where $(\bar{\ell},\bar{m})\neq(\ell,m)$. (\textit{Bottom Left and Right}) Factional residuals for mixing coefficient fits. For comparison, residual errors for the fits in Ref.~\cite{Berti:2014fga} are also shown over their region of validity in \bh{} spin. Note that makers match between grey and colored data points. }
  \label{fig:ys}
\end{figure*}

 \fig{fig:ys} displays fits, training data and related residuals.
For efficiency of presentation, each $\sigma_{\LMlmn}$ is plotted via its real and imaginary part.
%
In cases where $(\bar\ell,\bar{m})=(\ell,m)$, the real part of $\sigma_{\LMlmn}$ varies about unity in a manner consistent with the Schwarzschild limit, where $\sigma_{\LMlmn}=\delta_{\bar\ell}^{\ell}\;\delta_{\bar{m}}^{m}$.
Consistency with the Schwarzschild limit is equally true in cases where $(\bar\ell,\bar{m}) \neq (\ell,m)$.
As with the \qnm{} frequencies, residuals are dominated by small scale oscillations with amplitudes that are fractions of a percent of the central values, and largely result from the model asatz.
For comparison, fractional residual errors for models in Ref.~\cite{Berti:2014fga} are also shown.
\section{Discussion}
\label{discuss}
\par We have developed upon previous techniques for the linear and pseudo-linear modeling of low noise data.
In particular, the \gmvp{} algorithm performs multivariate polynomial modeling of real and complex valued scalar functions with no inherent limitation on the number of domain parameters.
The \gmvr{} algorithm does the same with multivariate rational functions.
When applied to the modeling of analytically computed quantities, both algorithms perform extremely well in producing accurate and precise representations of training data, suggesting extended applicability of \gmvr{} and \gmvp{} to similar problems.
\par Treating a toy problem with \gmvr{} demonstrates its ability to faithfully recover underlying model parameters for a plausible dataset.
This treatment also demonstrates the convergence of the algorithm's greedy phase with increasing iterations, as well as the convergence of an underlying iterative refinement phase (\ceqn{eq:rat4}).
\par Both \gmvp{} and \gmvr{} may be used to automatically determine the functional form and model for a given dataset that is expected to be respectively polynomial or rational.
An alternative use-strategy is to use either \gmvp{} or \gmvr{} to determine a fitting ansatz for individual cases (e.g. individual \qnm{s}), and then use these results to develop a single ansatz for all cases.
This is what as been done for the modeling of \qnm{} frequencies and spherical-spheroidal mixing coefficients.
\par \gmvp{} has been applied to the modeling of \qnm{} frequencies. The resulting models have been constrained in the extremal Kerr limit, and perform well compared to similar fits in the literature.
The fits presented here are of direct use in Ref. \cite{London:2017bcn}, where efficiently evaluable \qnm{} frequencies are required to generate template waveforms for \gw{} searches and parameter estimation.
The fits presented may find future use in Phenom or EOB based \gw{} models.
\par \gmvr{} has been applied to the modeling of mixing coefficients between the spherical and spheroidal harmonics.
%
{\scriptsize
\begin{algorithm}[H]
  \caption{\gmvr{}, a degree tempered stepwise algorithm for multivariate rational modeling of scalar data.}
  \label{alg:gmvr}
  \begin{algorithmic}[1]
    \State {\bf Input:} $ \{ x, f, $max\_degree = 6$, tol \}$
    \vskip 10pt
    \State Define, $\mlam_{bulk}$, the bulk symbol space, to be the set of all multinomial combinations of basis vectors up to a predefined maximum order. This is the combined symbol space for numerator and denominator symbols.
    \State Define $\mathcal{A}_{\gmvr{}}$ according to Alg. (\ref{alg:A_gmvr}).
    \State Given $max\_degree$, define, ${ D}$, a list of allowed polynomial degrees (e.g. $\{0,1,2,3,4,5,6\}$)
    \For {$d$ \textbf{in}  ${ D}$ }
      \State Define $\mlam_{bulk}^{(d)}$ as all symbols from $\mlam_{bulk}$ with degree less than or equal to current degree: $\mlam^{(d)}_{bulk}$
      \State Using $\mlam_{bulk}^{(d)}$, apply Alg. (\ref{alg:pgreedy}), \pgreedy{}, with $\mathcal{A}_{\gmvp{}}$ to get symbol subset, $\mlam^{(d)}_{opt}$ and estimator val, $\epsilon^{(d)}_{opt}$
      \If { $|\epsilon^{(d)}_{opt}-\epsilon^{(d-1)}_{opt}|<tol$ }
        \State \textbf{break}
      \EndIf
    \EndFor
    \vskip 10pt
    \State {\bf Output:} $\mlam^{(d)}_{opt}$
  \end{algorithmic}
\end{algorithm}
}
These fits are of direct use in Ref. \cite{Carullo:2018sfu}, and may be of future use in similar ringdown-only models for the purpose of testing \gr{}.
\par While \gmvr{} and \gmvp{} show promise in the cases shown here, in their presented rudimentary form, both posses a number of limitations.
If given sufficiently dense training data, neither currently performs cross-validation.
And perhaps most notably, neither method directly accounts for information about the noise distribution within the training data.
As such, the methods presented are recommended primarily for datasets where noise is very small or negligible.
Nevertheless, the \gmvr{} toy problem demonstrates \gmvr{}'s ability to handle moderately noisy training data, suggesting current applicability to a variety of problems where polynomial interpolation is insufficient.
\par Of relevance to current and future \gw{} science, the models presented for \qnm{} frequencies and harmonic mixing coefficients have aided (e.g. Refs. \cite{Carullo:2018sfu,London:2018gaq,London:2017bcn} ), and are expected to continue aiding the development and implementation of \gw{} signal models.

\section*{Acknowledgements}
%
The authors thank Mark Hannam for useful discussions.
The work presented in this paper was supported by Science and Technology Facilities Council (STFC)
grant ST/L000962/1, and European Research Council Consolidator Grant 647839.

\newpage

\appendix

\newpage
\section{Additional Equations}
\label{app:eqns}

%
%
\begin{widetext}
  \input{ysprod_eqns}
\end{widetext}



\newpage
\bibliographystyle{mnras}
\bibliography{mvf.bib}
\end{document}

%% file: definitions.tex
\usepackage{inputenc}
\usepackage{pgfplots}
\usepackage{tikz}
\usetikzlibrary{shapes,arrows}
\usetikzlibrary{positioning}
\usetikzlibrary{backgrounds}

\newcommand\red[1]{{\color[rgb]{0.75,0.0,0.0} #1}}

\newcommand{\mlam}{{\bm{\lambda}}}

\definecolor{lightblue}{rgb}{.82,.88,0.95}
\definecolor{lightred}{rgb}{0.95,.86,0.86}
\definecolor{yellow}{rgb}{0.95,0.95,0.86}
\definecolor{green}{rgb}{.90,1,0.95}
\definecolor{lightpurple}{rgb}{.95,0.85,0.95}

\tikzset{
    vertex/.style = {
        circle,
        fill            = black,
        outer sep = 2pt,
        inner sep = 1pt,
    }
}

\def\tk{Teukolsky}

\def\ee{Einstein's equations}

\def\grad#1{gravitational radiation#1}
\def\gr#1{General Relativity#1}

\def\gw#1{gravitational wave#1}

\def\nht#1{No-Hair Theorem#1}

\def\sec#1{Section~(\ref{#1})}

\def\tk#1{Teukolsky#1}

\def\fig#1{Fig.~\ref{#1}} 

\def\eqn#1{Eqn.~(\ref{#1})}
\def\ceqn#1{Eqn.~\ref{#1}}
\newcommand{\eqns}[2]{Equations~(\ref{#1})--(\ref{#2})}

\def\tbl#1{Table~(\ref{#1})}

\def\da#1{data analysis#1}
\def\imr#1{inspiral, merger and ringdown#1
  (IMR#1)\gdef\imr{IMR}}
\def\lal#1{LIGO Analysis Library#1
  (LAL#1)\gdef\lal{LAL}}
\def\nrda#1{\nr{} Data Analysis#1
  (NRDA#1)\gdef\nrda{NRDA}}
\def\tt#1{\textit{transverse--traceless}#1
  (TT#1)\gdef\tt{TT}}
\def\et#1{Einstein Telescope#1
  (ET#1)\gdef\et{ET}}
\def\ego#1{European Gravitational Observatory#1
  (EGO#1)\gdef\ego{EGO}}
\def\elisa#1{Evolved Laser Interferometer Space Antenna#1
  (eLISA#1)\gdef\elisa{eLISA}}
\def\ligo#1{Laser Interferometer Gravitational Wave Observatory#1
  (LIGO#1)\gdef\ligo{LIGO}}
\def\virgo#1{Virgo#1}
\def\aligo#1{Advanced LIGO#1
  (aLIGO#1)\gdef\aligo{aLIGO}}
\def\snr#1{signal-to-noise ratio#1
  (SNR#1)\gdef\snr{SNR}}
\def\psd#1{power spectral density#1
  (PSD#1)\gdef\psd{PSD}}
\def\rom#1{reduced order model#1
  (ROM#1)\gdef\rom{ROM}}
\def\gatech#1{Georgia Institute of Technology#1
  (GaTech#1)\gdef\gatech{GaTech}}
\def\ffi#1{Fixed-Frequrncy Integration#1
  (FFI#1)\gdef\ffi{FFI}}
\def\sxs#1{Simulating Extreme Spacetimes#1
  (SXS#1)\gdef\sxs{SXS}}
\def\bam#1{Bifunctional Adaptive Mesh#1
  (BAM#1)\gdef\bam{BAM}}
\def\adm#1{Arnowitt-Deser-Misner
	(ADM#1)\gdef\adm{ADM}}
\def\frmse#1{Fractional Root-Mean Square Error
	(FRMSE)\gdef\frmse{FRMSE}}

\def\bh#1{black hole#1
 (BH#1)\gdef\bh{BH}}
\def\bbh#1{binary \bh{}#1}

\def\qnm#1{Quasi-Normal Mode#1
  (QNM#1)\gdef\qnm{QNM}}
\def\eob#1{Effective One Body#1
  (EOB#1)\gdef\eob{EOB}}
\def\gw#1{gravitational-wave#1}
\def\gw#1{Gravitational wave#1
 (GW#1)\gdef\gw{GW}}

\def\spa#1{stationary phase approximation#1
 (SPA#1)\gdef\spa{SPA}}
\def\pn#1{Post-Newtonian#1
	(PN#1)\gdef\pn{PN}}
\def\pnl#1{post-Newtonian-like#1
  (PN-like#1)\gdef\pnl{PN-like}}

\def\nr{Numerical Relativity
 (NR)\gdef\nr{NR}}

\def\rd{ringdown}

\def\pca#1{principle component analysis#1
  (PCA#1)\gdef\pca{PCA}}
\def\svd#1{Singular Value Decomposition#1
  (SVD#1)\gdef\svd{SVD}}

\newcommand\hidetosubmit[1]{}
\newcommand\optional[1]{}
\newcommand\ForInternalReference[1]{}

%% file: cwfit_eqns.tex
\begin{align}
	\label{eq:cw_fit_1}
	\cw_{220}(\kappa) \; &= \;\, 1.0 \, + \, \kappa \, (1.5578 e^{2.9031 i}\, + \, 1.9510  e^{5.9210 i} \kappa\, + \, 2.0997  e^{2.7606 i} \kappa ^ 2\, + \, 1.4109  e^{5.9143 i} \kappa ^ 3\, + \, 0.4106  e^{2.7952 i} \kappa ^ 4 \, )  \\
	\label{eq:cw_fit_2}
	\cw_{221}(\kappa) \; &= \;\, 1.0 \, + \, \kappa \, (1.8709 e^{2.5112 i}\, + \, 2.7192  e^{5.4250 i} \kappa\, + \, 3.0565  e^{2.2857 i} \kappa ^ 2\, + \, 2.0531  e^{5.4862 i} \kappa ^ 3\, + \, 0.5955  e^{2.4225 i} \kappa ^ 4 \, )  \\
	\label{eq:cw_fit_3}
	\cw_{330}(\kappa) \; &= \;\, 1.5 \, + \, \kappa \, (2.0957 e^{2.9650 i}\, + \, 2.4696  e^{5.9967 i} \kappa\, + \, 2.6655  e^{2.8176 i} \kappa ^ 2\, + \, 1.7584  e^{5.9327 i} \kappa ^ 3\, + \, 0.4991  e^{2.7817 i} \kappa ^ 4 \, )  \\ 
	\label{eq:cw_fit_4}
	\cw_{331}(\kappa) \; &= \;\, 1.5 \, + \, \kappa \, (2.3391 e^{2.6497 i}\, + \, 3.1399  e^{5.5525 i} \kappa\, + \, 3.5916  e^{2.3472 i} \kappa ^ 2\, + \, 2.4490  e^{5.4435 i} \kappa ^ 3\, + \, 0.7004  e^{2.2830 i} \kappa ^ 4 \, )  \\
	\label{eq:cw_fit_5}
	\cw_{440}(\kappa) \; &= \;\, 2.0 \, + \, \kappa \, (2.6589 e^{3.0028 i}\, + \, 2.9783  e^{6.0510 i} \kappa\, + \, 3.2184  e^{2.8775 i} \kappa ^ 2\, + \, 2.1276  e^{5.9897 i} \kappa ^ 3\, + \, 0.6034  e^{2.8300 i} \kappa ^ 4 \, )  \\
	\label{eq:cw_fit_6}
	\cw_{430}(\kappa) \; &= \;\, 1.5 \, + \, \kappa \, (0.2050 e^{0.5953 i}\, + \, 3.1033  e^{3.0162 i} \kappa\, + \, 4.2361  e^{6.0388 i} \kappa ^ 2\, + \, 3.0289  e^{2.8262 i} \kappa ^ 3\, + \, 0.9084  e^{5.9152 i} \kappa ^ 4 \, )  \\
	\label{eq:cw_fit_7}
	\cw_{550}(\kappa) \; &= \;\, 2.5 \, + \, \kappa \, (3.2405 e^{3.0279 i}\, + \, 3.4906  e^{6.0888 i} \kappa\, + \, 3.7470  e^{2.9212 i} \kappa ^ 2\, + \, 2.4725  e^{6.0365 i} \kappa ^ 3\, + \, 0.6994  e^{2.8766 i} \kappa ^ 4 \, )  \\
	\label{eq:cw_fit_8}
	\cw_{320}(\kappa) \; &= \;\,1.0225 e^{0.0049 i}\, + \, 0.2473  e^{0.6653 i} \kappa\, + \, 1.7047  e^{3.1383 i} \kappa ^ 2\, + \, 0.9460  e^{0.1632 i} \kappa ^ 3\, + \, 1.5319  e^{5.7036 i} \kappa ^ 4\\ \nonumber
	&\hspace{255pt}\, + \, 2.2805  e^{2.6852 i} \kappa ^ 5\, + \, 0.9215  e^{5.8417 i} \kappa ^ 6 \\
	\label{eq:cw_fit_9}
	\cw_{210}(\kappa) \; &= \;\,0.5891 e^{0.0435 i}\, + \, 0.1890  e^{2.2899 i} \kappa\, + \, 1.1501  e^{5.8101 i} \kappa ^ 2\, + \, 6.0459  e^{2.7420 i} \kappa ^ 3\, + \, 11.1263  e^{5.8441 i} \kappa ^ 4\\ \nonumber
	&\hspace{255pt}\, + \, 9.3471  e^{2.6694 i} \kappa ^ 5\, + \, 3.0384  e^{5.7915 i} \kappa ^ 6
\end{align}

%% file: ysprod_eqns.tex
\begin{align}
  \label{eq:ys22220}
  \sigma_{22220} \, &= \, 0.99733\,e^{6.2813i} \, + 0.0075336 \, \frac{   14.592\,e^{5.0601i}\,\kappa \, + \, (28.761\,e^{1.629i})\,{\kappa}^{2} \, + \, (14.511\,e^{4.6362i})\,{\kappa}^{3} \, + \, (1.9624\,e^{3.0113i})  }{ 1 \, + \,   0.88674\,e^{3.0787i}\,\kappa \, + \, (1.002\,e^{0.13211i})\,{\kappa}^{2} \, + \, (0.082148\,e^{5.6369i})\,{\kappa}^{3} }   \\
  \label{eq:ys21210}
  \sigma_{21210} \, &= \, 0.99716\,e^{6.2815i} \, + 0.0063542 \, \frac{   1.4345\times 10^5\,e^{4.5061i}\,\kappa \, + \, (3.5469\times 10^5\,e^{1.7327i})\,{\kappa}^{2} \, + \, (2.4038\times 10^5\,e^{5.1629i})\,{\kappa}^{3} \, + \, (6026.9\,e^{1.8881i})  }{ 1 \, + \,   73780\,e^{4.5545i}\,\kappa \, + \, (97494\,e^{1.398i})\,{\kappa}^{2} \, + \, (34815\,e^{4.5623i})\,{\kappa}^{3} }   \\
  \label{eq:ys22221}
  \sigma_{22221} \, &= \, 0.99683\,e^{6.2782i} \, + 0.020758 \, \frac{   15.077\,e^{4.8323i}\,\kappa \, + \, (31.139\,e^{1.585i})\,{\kappa}^{2} \, + \, (15.449\,e^{4.6727i})\,{\kappa}^{3} \, + \, (0.71897\,e^{2.8084i})  }{ 1 \, + \,   0.80592\,e^{3.3995i}\,\kappa \, + \, (0.69502\,e^{0.54275i})\,{\kappa}^{2} \, + \, (0.35613\,e^{5.9545i})\,{\kappa}^{3} }   \\
  \label{eq:ys32320}
  \sigma_{32320} \, &= \, 0.99009\,e^{6.2804i} \, + 0.02369 \, \frac{   71893\,e^{1.2395i}\,\kappa \, + \, (1.7055\times 10^5\,e^{5.0371i})\,{\kappa}^{2} \, + \, (1.2947\times 10^5\,e^{2.359i})\,{\kappa}^{3} \, + \, (1935.5\,e^{4.668i})  }{ 1 \, + \,   38206\,e^{1.2254i}\,\kappa \, + \, (35811\,e^{3.9618i})\,{\kappa}^{2} \, + \, (8378.3\,e^{0.11726i})\,{\kappa}^{3} }   \\
  \label{eq:ys33331}
  \sigma_{33331} \, &= \, 0.99478\,e^{6.2688i} \, + 0.040478 \, \frac{   4.4113\,e^{1.2501i}\,\kappa \, + \, (11.588\,e^{0.27959i})\,{\kappa}^{2} \, + \, (17.322\,e^{3.7904i})\,{\kappa}^{3} \, + \, (0.67724\,e^{2.5797i})  }{ 1 \, + \,   3.8782\,e^{2.2864i}\,\kappa \, + \, (3.4913\,e^{5.6655i})\,{\kappa}^{2} \, + \, (1.0368\,e^{2.9082i})\,{\kappa}^{3} }   \\
  \label{eq:ys32221}
  \sigma_{32221} \, &= \, 0.02203\,e^{0.16452i} \, + 0.073233 \, \frac{   24.932\,e^{1.0181i}\,\kappa \, + \, (30.197\,e^{4.4047i})\,{\kappa}^{2} \, + \, (11.274\,e^{2.981i})\,{\kappa}^{3} \, + \, (2.4374\,e^{6.1959i})  }{ 1 \, + \,   11.397\,e^{3.9953i}\,\kappa \, + \, (10.915\,e^{5.8025i})\,{\kappa}^{2} \, + \, (7.2196\,e^{1.8176i})\,{\kappa}^{3} }   \\
  \label{eq:ys33330}
  \sigma_{33330} \, &= \, 0.99569\,e^{6.2785i} \, + 0.014546 \, \frac{   7.2112\,e^{0.62811i}\,\kappa \, + \, (6.5381\,e^{4.6216i})\,{\kappa}^{2} \, + \, (4.451\,e^{2.9228i})\,{\kappa}^{3} \, + \, (1.7113\,e^{2.9527i})  }{ 1 \, + \,   1.4974\,e^{1.6687i}\,\kappa \, + \, (1.5288\,e^{5.3885i})\,{\kappa}^{2} \, + \, (0.52114\,e^{2.5471i})\,{\kappa}^{3} }   \\
  \label{eq:ys32220}
  \sigma_{32220} \, &= \, 0.020598\,e^{0.04743i} \, + 0.06919 \, \frac{   2.7657\,e^{2.133i}\,\kappa \, + \, (3.9562\,e^{4.653i})\,{\kappa}^{2} \, + \, (2.3364\,e^{2.6444i})\,{\kappa}^{3} \, + \, (2.399\,e^{6.2767i})  }{ 1 \, + \,   1.0595\,e^{4.7865i}\,\kappa \, + \, (0.91308\,e^{2.887i})\,{\kappa}^{2} \, + \, (0.69468\,e^{0.1912i})\,{\kappa}^{3} }   \\
  \label{eq:ys43330}
  \sigma_{43330} \, &= \, 0.028112\,e^{0.048488i} \, + 0.086383 \, \frac{   12.087\,e^{0.47221i}\,\kappa \, + \, (30.626\,e^{3.3281i})\,{\kappa}^{2} \, + \, (16.328\,e^{6.1785i})\,{\kappa}^{3} \, + \, (2.3603\,e^{6.2662i})  }{ 1 \, + \,   4.9638\,e^{3.5931i}\,\kappa \, + \, (6.2552\,e^{6.2001i})\,{\kappa}^{2} \, + \, (1.4538\,e^{2.5539i})\,{\kappa}^{3} }   \\
  \label{eq:ys43430}
  \sigma_{43430} \, &= \, 0.98735\,e^{6.2795i} \, + 0.033028 \, \frac{   7.0084\times 10^5\,e^{1.1067i}\,\kappa \, + \, (1.843\times 10^6\,e^{4.8808i})\,{\kappa}^{2} \, + \, (1.4367\times 10^6\,e^{2.1412i})\,{\kappa}^{3} \, + \, (13844\,e^{4.5601i})  }{ 1 \, + \,   3.5667\times 10^5\,e^{1.0149i}\,\kappa \, + \, (3.274\times 10^5\,e^{3.7746i})\,{\kappa}^{2} \, + \, (88621\,e^{0.10095i})\,{\kappa}^{3} }   \\
  \label{eq:ys44440}
  \sigma_{44440} \, &= \, 0.99478\,e^{6.2776i} \, + 0.024791 \, \frac{   6.5172\,e^{0.79835i}\,\kappa \, + \, (7.7748\,e^{4.2485i})\,{\kappa}^{2} \, + \, (1.1577\,e^{1.5905i})\,{\kappa}^{3} \, + \, (1.2434\,e^{2.9616i})  }{ 1 \, + \,   0.44548\,e^{1.2496i}\,\kappa \, + \, (0.59437\,e^{5.6732i})\,{\kappa}^{2} \, + \, (0.24743\,e^{2.8292i})\,{\kappa}^{3} }   \\
  \label{eq:ys55550}
  \sigma_{55550} \, &= \, 0.99434\,e^{6.2773i} \, + 0.03126 \, \frac{   6.5508\,e^{0.93398i}\,\kappa \, + \, (8.0558\,e^{4.2881i})\,{\kappa}^{2} \, + \, (0.92971\,e^{1.0436i})\,{\kappa}^{3} \, + \, (1.0904\,e^{2.9712i})  }{ 1 \, + \,   0.23128\,e^{1.7666i}\,\kappa \, + \, (0.54958\,e^{5.9178i})\,{\kappa}^{2} \, + \, (0.213\,e^{3.0092i})\,{\kappa}^{3} }  
\end{align}